\documentclass[twocolumn,prb,preprintnumbers,amsmath,amssymb,showpacs]{revtex4}
\usepackage{lipsum} 
\usepackage{hyperref}
\usepackage[latin9]{inputenc}
\usepackage{graphicx}
\usepackage{dcolumn}
\usepackage{bbold}
\usepackage{bm}
\usepackage[usenames]{xcolor}

\makeatletter
\newsavebox\myboxA
\newsavebox\myboxB
\newlength\mylenA

\newcommand*\xoverline[2][0.75]{%
    \sbox{\myboxA}{$\m@th#2$}%
    \setbox\myboxB\null
    \ht\myboxB=\ht\myboxA%
    \dp\myboxB=\dp\myboxA%
    \wd\myboxB=#1\wd\myboxA
    \sbox\myboxB{$\m@th\overline{\copy\myboxB}$}
    \setlength\mylenA{\the\wd\myboxA}
    \addtolength\mylenA{-\the\wd\myboxB}%
    \ifdim\wd\myboxB<\wd\myboxA%
       \rlap{\hskip 0.5\mylenA\usebox\myboxB}{\usebox\myboxA}%
    \else
        \hskip -0.5\mylenA\rlap{\usebox\myboxA}{\hskip 0.5\mylenA\usebox\myboxB}%
    \fi}
\makeatother

\begin{document}

\title{Strong competition between $\Theta_{II}$-loop-current order and
$d$-wave charge order along the diagonal direction in a two-dimensional
hot spot model}

\author{Vanuildo S. de Carvalho$^{1,2}$}
\author{Thomas Kloss$^{2}$}
\author{Xavier Montiel$^{2}$} 
\author{Hermann Freire$^{1}$} 
\email{hermann\_freire@ufg.br}
\author{Catherine Pépin$^{2}$}

\affiliation{$^{1}$Instituto de Física, Universidade Federal de Goiás, 74.001-970,
Goiânia-GO, Brazil,}

\affiliation{$^{2}$IPhT, L'Orme des Merisiers, CEA-Saclay, 91191 Gif-sur-Yvette,
France}

\date{\today}
\begin{abstract}
We study the fate of the so-called $\Theta_{II}$-loop-current order
that breaks both time-reversal and parity symmetries in a two-dimensional
hot spot model with antiferromagnetically mediated interactions, using 
Fermi surfaces relevant to the phenomenology of the cuprate
superconductors. We start from a three-band Emery
model describing the hopping of holes in the CuO$_{2}$
plane that includes two hopping parameters $t_{pp}$ and $t_{pd}$,
local on-site Coulomb interactions $U_{d}$ and $U_{p}$ and nearest-neighbor
$V_{pd}$ couplings between the fermions in the copper {[}Cu$(3d_{x^{2}-y^{2}})${]}
and oxygen {[}O$(2p_{x})$ and O$(2p_{y})${]} orbitals. By focusing
on the lowest-energy band, we proceed to decouple the local interaction
$U_{d}$ of the Cu orbital in the spin channel using a Hubbard-Stratonovich
transformation to arrive at the interacting part of the so-called
spin-fermion model. We also decouple the nearest-neighbor interaction
$V_{pd}$ to introduce the order parameter of the $\Theta_{II}$-loop-current
order. In this way, we are able to construct a consistent mean-field
theory that describes the strong competition between the composite order
parameter made of a quadrupole-density-wave and $d$-wave pairing
fluctuations proposed in Efetov \emph{et al.} {[}Nat. Phys. \textbf{9},
442 (2013){]} with the $\Theta_{II}$-loop-current order parameter
that is argued to be relevant for explaining important aspects of
the physics of the pseudogap phase displayed in the underdoped cuprates.
\end{abstract}


\maketitle

\section{Introduction}

The physics of the pseudogap phase of cuprate superconductors
remains one of the most enduring open problems of condensed
matter physics. There are recent pervasive hints that the 
pseudogap phase in most underdoped cuprate superconductors
might involve one or more symmetry-breaking ``hidden'' orders,
whose precise microscopic mechanisms are still elusive to this date.
State-of-the-art experiments such as nuclear magnetic resonance \cite{Julien,Julien2},
pulsed-echo ultrasound experiments \cite{LeBoeuf}, x-ray scattering
\cite{Ghiringhelli,Achkar,Chang2} and scanning tunneling microscopy
\cite{Hoffman,Yazdani} performed in non-Lanthanum-based materials
established the emergence of a dome-shaped short-range incommensurate
$d$-wave \cite{Comin,Fujita} charge-density-wave (CDW) at low hole
doping with a modulation described by the wavevectors $\mathbf{Q_{x}}=(Q_{0},0)$
and $\mathbf{Q_{y}}=(0,Q_{0})$ oriented along the principal
axes of the CuO$_{2}$ unit cell (with $Q_{0}\simeq0.255$ in reciprocal
lattice units \cite{Comin2,SilvaNeto}). Quite surprisingly, the peak
of this short-range charge order dome occurs approximately at the
universal hole doping $x\simeq0.12$ for several compounds \cite{LouisTaillefer},
despite their differences in material-specific properties. This could
suggest that simple, low-energy effective models may potentially capture
the essence of the physics of these materials \cite{Chubukov,Metlitski,Efetov,LaPlaca}.
We will follow this point of view in the present work. Moreover, by
applying pressure on these systems, the charge order can be completely
suppressed, while the pseudogap phase remains unaffected \cite{LouisTaillefer}.
This clearly indicates that such a CDW order emerges on top of an
already-formed pseudogap phase, instead of being its driving force.
On the other hand, at very high magnetic fields, the $d$-wave superconducting
phase displayed by these materials is destroyed and the short-range
CDW turns into a long-range order. In this context, 
it plays a central role in reconstructing the Fermi surface of these
compounds into pockets, as is evidenced in quantum oscillation experiments
\cite{Taillefer,Sebastian}.

In addition to these salient features taking place within the pseudogap
phase, another form of ``hidden'' order representing potentially one
of the driving forces of quantum criticality in these systems (that
may even coexist with CDW order and $d$-wave superconductivity at
lower temperature scales $T<T^{*}$) is suggested by a different set
of equally groundbreaking experiments: spin-polarized neutron scattering
\cite{Greven,Bourges_2} and Kerr-rotation experiments \cite{Kapitulnik,Kapitulnik_2}
indicate spontaneous breaking of both time-reversal and parity symmetries
in this phase at temperatures that are reasonably close to $T^{*}$
over a wide doping range. This phase transition has thus been referred
to as the Kerr transition in the literature. This transition was
given a theoretical framework in the proposal by Varma \cite{Varma} (see also an interesting, alternative proposal put forward in Ref.  \cite{Bulut})
that orbital loop current order -- we shall specialize in the present work to the so-called
$\Theta_{II}$-phase -- may account for the observed properties in these
materials, since it naturally preserves the translational symmetry
of the lattice and, additionally, it leads to the breaking of the
correct discrete symmetries consistent with spin
polarized neutron scattering experiments \cite{Fauque}. This
theoretical description requires starting from at least
a three-band model, in which one includes besides the usually considered
copper $d_{x^{2}-y^{2}}$-orbital, also the oxygen $p_{x}$ and $p_{y}$-orbitals
of the CuO$_{2}$ unit cell. Such a minimal model turns out to be
essential to describe intra-unit-cell loop currents involving charge
transfer between oxygen orbitals that appear in the aforementioned
$\Theta_{II}$-phase. This theoretical proposal is physically appealing
but it has one potential disagreement with experiments: it is hard
to obtain the result that the underlying Fermi surface gaps out at
all, since the phase transition does not break translational symmetry near
the hot spots (i.e. the points in momentum space where the Fermi surface
intersects the antiferromagnetic Brillouin zone boundary). Moreover,
it is important to mention that recently a quantum critical point
(QCP) was revealed in the cuprates at a hole-doping $x_{crit}\simeq0.18$
via an analysis of the quasiparticle mass enhancement using quantum
oscillation experiments \cite{Harrison}. Interestingly, this critical
point may represent approximately the termination of the Kerr transition
line, the charge-order-dome and an as-yet-unidentified third phase
competing with the previous two orders at a doping level reasonably
close to optimal doping.

On the theoretical front, the hot spot model emerges as an interesting,
minimal low-energy effective model that captures qualitative aspects
of the physics of the high-$T_{c}$ cuprates from a weak-to-moderate
coupling perspective. In this respect, an important work by Metlitski
and Sachdev \cite{Metlitski} consisted in the elegant demonstration
that, if the energy dispersion of this model is linearized, an exact
emergent $SU(2)$ pseudospin symmetry relating a $d$-wave singlet
superconducting (SSC) order to a $d$-wave quadrupole-density-wave
(QDW) order at wavevectors along the Brillouin zone diagonal $(\pm Q_{0},\pm Q_{0})$
is verified at the spin-density-wave (SDW) quantum critical point.
This degeneracy between these two orders effectively produces a composite
order parameter (denoted by QDW/SSC) with both bond order and preformed
pairs at high temperatures as shown by Efetov \emph{et al.} \cite{Efetov}
and the properties of this state have been explored in connection
with the physics of the cuprates using different approaches in many
works \cite{Efetov_3,Efetov_2,Chowdhury,Sau,Wang,Allais,Allais_2,Tsvelik,Freire5,Freire7}.
In addition to this fact, another emergent $SU(2)$ degeneracy relating
two additional orders -- a superconducting order with a finite Cooper-pair
center of mass momentum (the so-called pair-density-wave (PDW) \cite{PALee,Agterberg,Fradkin})
and a $d$-wave CDW at the experimentally observed wavevectors $\mathbf{Q_{x}}$
and $\mathbf{Q_{y}}$ -- has also been recently verified in the model
in the work by Pépin \emph{et al.}\cite{Pepin_2} and explored further
by Wang \emph{et al.}\cite{Wang_2}. This additional degeneracy generates
another composite order parameter (denoted by PDW/CDW) with similar
energy scales that also competes with the QDW/SSC order \cite{Pepin_2,Wang_2,Freire6}.

In the present work, we will consider the relevant scenario in which
yet another order parameter (the $\Theta_{II}$-loop current order)
competes with the QDW/SSC order in an effective hot spot model. The
purpose of this study is to demonstrate the possibility that, due
to this competition, QSW/SSC is strongly affected by the $\Theta_{II}$-order
parameter that breaks both time-reversal and parity symmetries, but
instead preserves their product. This opens an interesting avenue
for future research and could be an explanation as to why the charge-order
signal in the cuprates is always observed along the axial vectors
(i.e. $\mathbf{Q_{x}}$ and $\mathbf{Q_{y}})$ and never along the
diagonal direction. In order to perform this investigation, we will
construct a novel mean-field theory by including both $\Theta_{II}$-loop
current order and the QDW/SSC composite order parameter in such an
effective model. As will become clear shortly, we will confirm in
this analysis the strong competition between $\Theta_{II}$-loop current
order and the QDW/SSC entangled order, with one order parameter being
clearly always detrimental to the other. Then we proceed to discuss
the physical implications of this strong competition for the physics
of the underdoped cuprates, in light of the recent experiments performed
in these materials.

Technically speaking, we will introduce a three-band model (Emery
model) describing hopping of holes in the CuO$_{2}$ plane which includes
two hopping parameters $t_{pp}$ and $t_{pd}$, on-site $U_{d}$ and
$U_{p}$ local interactions and nearest-neighbor $V_{pd}$ couplings
between the fermions in the copper ($d_{x^{2}-y^{2}}$) and oxygen
($p_{x}$ and $p_{y}$) orbitals. By focussing on the lowest-energy
band, we will decouple the local interaction $U_{d}$ of the Cu orbital
in the spin channel using a conventional Hubbard-Stratonovich transformation
to arrive at the interacting part of the so-called spin-fermion model.
Then, we will follow closely the methodology explained in full detail
in the paper by Efetov \emph{et al.} \cite{Efetov} to define the
composite order parameter associated with the QDW/SSC fluctuations.
In addition to this, we will also decouple the nearest-neighbor interaction
$V_{pd}$ of the model to introduce the order parameter associated
with the $\Theta_{II}$-loop-current order. Lastly, we will proceed
to derive analytically and then solve numerically the resulting mean-field
equations, which describes the competition between these two order
parameters.

This paper is organized as follows. In Section II, we define the three-band
model that we will be interested in and we show how to decouple the
interactions to obtain the resulting mean-field equations describing
the competition between the two orders. Since the interactions that
promote QDW/SSC and $\Theta_{II}$-loop current order turn out to
be different, this decoupling is unambiguous. In Section III, we solve
numerically the self-consistent mean-field equations and then we discuss
our main results. Finally, Section IV is devoted to our conclusions.

\section{The Three-band model}

We start this section by writing down both the noninteracting and
interacting Hamiltonians of the so-called three-band (Emery) model
following Refs. \cite{Emery,Varma_2,Hanke,Fischer} in order to describe the underdoped
cuprates as follows

\begin{align}
\mathcal{H}_{0}& = -t_{pd}\sum_{i,\sigma}\sum_{\nu}(\hat{d}_{i,\sigma}^{\dagger}\hat{p}_{i+\hat{\nu}/2,\sigma}+H.c.)\nonumber \\
 & -t_{pp}\sum_{i,\sigma}\sum_{\langle\nu,\nu'\rangle}(\hat{p}_{i+\hat{\nu}/2,\sigma}^{\dagger}\hat{p}_{i+\hat{\nu}'/2,\sigma}+H.c.)\nonumber \\
 & +(\varepsilon_{d}-\mu)\sum_{i,\sigma}\hat{n}_{i,\sigma}^{d}+\frac{1}{2}(\varepsilon_{p}-\mu)\sum_{i,\sigma}\sum_{\nu}\hat{n}_{i+\hat{\nu}/2,\sigma}^{p},\label{Eq_01}
\end{align}

\begin{align}
\mathcal{H}_{int}& = U_{d}\sum_{i}\hat{n}_{i,\uparrow}^{d}\hat{n}_{i,\downarrow}^{d}+\frac{U_{p}}{2}\sum_{i,\nu}\hat{n}_{i+\hat{\nu}/2,\uparrow}^{p}\hat{n}_{i+\hat{\nu}/2,\downarrow}^{p}\nonumber \\
 & +V_{pd}\sum_{i,\nu}\sum_{\sigma,\sigma'}\hat{n}_{i,\sigma}^{d}\hat{n}_{i+\hat{\nu}/2,\sigma'}^{p}.\label{Eq_02}
\end{align}

\begin{figure}[t]
\begin{centering}
\includegraphics[width=0.8 \linewidth]{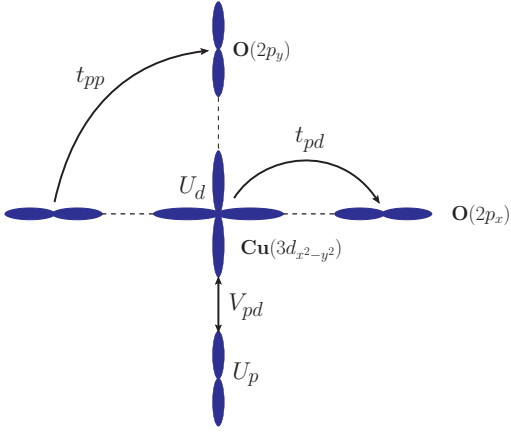}
\caption{(Color online) Orbital structure and the interactions of the three-band
model in the CuO$_{2}$ unit cell.}\label{Three_Band_Model_01}
\end{centering}
\end{figure}

\noindent This model Hamiltonian describes the fermionic motion on
the copper {[}Cu$(3d_{x^{2}-y^{2}})${]} and oxygen {[}O$(2p_{x})$
and O$(2p_{y})${]} orbitals that are located in the CuO$_{2}$ unit
cell (see Fig. \ref{Three_Band_Model_01}). The quantities $\hat{d}_{i,\sigma}^{\dagger}$, $\hat{d}_{i,\sigma}$, $\hat{p}_{i+\hat{\nu}/2,\sigma}^{\dagger}$
and $\hat{p}_{i+\hat{\nu}/2,\sigma}$ are, respectively, the creation and annihilation
operators of fermions situated on the site $i$ with spin $\sigma$
of the Cu orbital and the
the creation and annihilation operators of fermions on the site $i+\hat{\nu}/2$
($\nu=x,y$) with spin $\sigma$ of the O orbitals. Besides, $\hat{n}_{i,\sigma}^{d}$ and $\hat{n}_{i+\hat{\nu}/2,\sigma}^{p}$
correspond, respectively, to the fermionic number operators for particles located on the Cu  and O orbitals. The model also takes
into account pair hopping ($t_{pd}$ and $t_{pp}$), on-site ($U_{d}$
and $U_{p}$) and nearest-neighbor ($V_{pd}$) interactions involving
the fermions on the Cu and O orbitals. The parameters $\varepsilon_{d}$ and $\varepsilon_{p}$ are, respectively, the Cu and O orbital energies and $\mu$ is the chemical potential 
which controls the electronic density in the system.

Following Abanov and Chubukov \cite{Chubukov}, we first decouple
the $U_{d}$ part of the interacting Hamiltonian in the spin channel
using a conventional Hubbard-Stratonovich transformation. The resulting
action becomes 
\begin{align}\label{Eq_03}
 & \mathcal{S}^{^{(1)}}_{int}[d,\vec{\phi}]\nonumber \\
 & =\lambda\int d\tau\sum_{i}d_{i,\sigma}^{\dagger}\vec{\phi}_{i}\cdot\vec{\sigma}_{\sigma,\sigma'}d_{i,\sigma'}e^{i\mathbf{Q}\cdot\mathbf{r}_{i}}\nonumber \\
 & +\frac{1}{2}\int d\tau d^{2}\mathbf{r}\left[\frac{1}{v_{s}^{2}}(\partial_{\tau}\vec{\phi})^{2}+(\nabla\vec{\phi})^{2}+m_{a}\vec{\phi}^{2}+\frac{g}{2}((\vec{\phi})^{2})^{2}\right],
\end{align}
where the bosonic field $\vec{\phi}_{i}=(\phi_{i}^{x},\phi_{i}^{y},\phi_{i}^{z})$
is the spin-density wave (SDW) order parameter at the antiferromagnetic
wave vector $\mathbf{Q}=(\pi,\pi)$, $v_{s}$ is the spin-wave velocity,
and $m_{a}$ is the spin-wave bosonic mass which vanishes at the quantum
critical point (QCP) of the theory. The $\sigma^{a}$ $(a=x,y,z)$
are the usual Pauli matrices. Notice that in Eq. \eqref{Eq_03} we
have partially integrated out the high-energy fermions in order to
derive an effective theory $\mathcal{S}^{^{(1)}}_{int}[d,\vec{\phi}]$ that
corresponds to the so-called spin-fermion model describing the coupling
between the itinerant low-energy fermionic excitations and the antiferromagnetic
SDW fluctuations. Another possibility in order to investigate the Emery model is to start from a more localized picture by mapping the
model defined in Eqs. (1) and (2) onto an effective three-band $t-J$ model \cite{Thomale}. We, however, will not
follow this latter route in the present work. For this reason, we would like to state clearly from the outset that our starting point here will be a more itinerant picture.

\begin{figure}[b]
\begin{center}
\includegraphics[width=0.9 \linewidth]{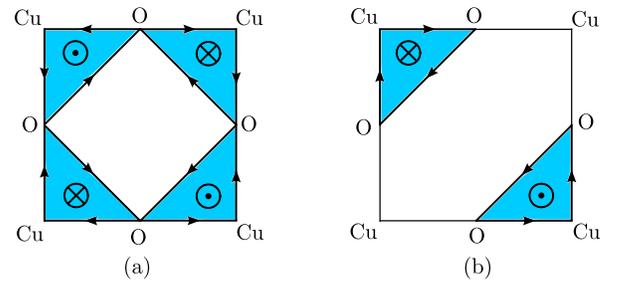}
\caption{(Color online) Loop current pattern in the CuO$_{2}$ unit cell for
the $\Theta_{I}$- and $\Theta_{II}$-loop current phases {[}panels
(a) and (b), respectively{]} proposed by Varma to explain the
physical properties of the pseudogap state in high-$T_{c}$ cuprate
superconductors. The symbols ($\odot$) and ($\otimes$) represent
the orientation of the local magnetic moments generated by the loop currents.}\label{Three_Band_Model_02}
\end{center}
\end{figure}

\begin{figure}[h]
\begin{center}
\includegraphics[width=0.6 \linewidth]{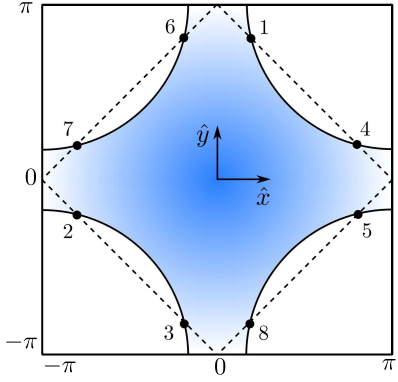}
\caption{(Color online) Representation of the Brillouin zone with the underlying
noninteracting Fermi surface (that encloses the blue area) which characterizes
the underdoped cuprate superconductors. The small black circles denote
the so-called hot spots which are defined as the intersection of the
Fermi surface with the antiferromagnetic zone boundary. For instance,
the hot spot labeled as $1$ has a wavevector $\mathbf{k_{1}}=(K_{-},K_{+})$
in momentum space with the constraint $K_{-}+K_{+}=\pi$. The wavevectors
of all the other hot spots in the Brillouin zone are obtained by simple
symmetry operations.}\label{Three_Band_Model_03}
\end{center}
\end{figure}

Now we turn our attention to the $V_{pd}$ interaction term in Eq.
\eqref{Eq_02}. This can be rewritten as 
\begin{equation}
V_{pd}\sum_{i,\nu}\sum_{\sigma,\sigma'}\hat{n}_{i,\sigma}^{d}\hat{n}_{i+\hat{\nu}/2,\sigma'}^{p}=-V_{pd}\sum_{i,j}\sum_{\sigma,\sigma'}\mathcal{A}_{i,\sigma}^{\dagger(j)}\mathcal{A}_{i,\sigma'}^{(j)},\label{Eq_04}
\end{equation}
with the field operators on the right-hand-side of the above equality
being 
\begin{align}
\mathcal{A}_{i,\sigma}^{\dagger(1,2)} & =\frac{1}{2}[(\hat{d}_{i,\sigma}^{\dagger}\hat{p}_{i+\hat{x}/2,\sigma}+\hat{d}_{i,\sigma}^{\dagger}\hat{p}_{i-\hat{x}/2,\sigma})\nonumber \\
 & \pm(\hat{d}_{i,\sigma}^{\dagger}\hat{p}_{i+\hat{y}/2,\sigma}+\hat{d}_{i,\sigma}^{\dagger}\hat{p}_{i-\hat{y}/2,\sigma})],\label{Eq_05}\\
\mathcal{A}_{i,\sigma}^{\dagger(3,4)} & =\frac{i}{2}[(\hat{d}_{i,\sigma}^{\dagger}\hat{p}_{i+\hat{x}/2,\sigma}-\hat{d}_{i,\sigma}^{\dagger}\hat{p}_{i-\hat{x}/2,\sigma})\nonumber \\
 & \pm(\hat{d}_{i,\sigma}^{\dagger}\hat{p}_{i+\hat{y}/2,\sigma}-\hat{d}_{i,\sigma}^{\dagger}\hat{p}_{i-\hat{y}/2,\sigma})].\label{Eq_06}
\end{align}
As first shown by Varma\cite{Varma}, only the order parameters associated
with $\mathcal{A}_{i,\sigma}^{(2)}$, $\mathcal{A}_{i,\sigma}^{(3)}$,
and $\mathcal{A}_{i,\sigma}^{(4)}$ lead to states with the presence
of stationary-loop currents on the CuO$_{2}$ plane and, of course,
to time-reversal symmetry breaking. The loop-current order with order
parameter defined in terms of $\mathcal{A}_{i,\sigma}^{(2)}$ is conventionally
called the $\Theta_{I}$-loop current phase, while the loop-current
order with order parameter given in terms of $\mathcal{A}_{i,\sigma}^{(3)}$,
and $\mathcal{A}_{i,\sigma}^{(4)}$ are known as the $\Theta_{II}$-loop
current phase (see Fig. \ref{Three_Band_Model_02}). In view of the
interpretation of some experiments on the pseudogap phase of the cuprate
superconductors as an evidence in favor of the $\Theta_{II}$-loop
current phase\cite{Fauque,Greven}, we will analyze henceforth only
this type of order. In this way, the decoupling of the interacting
term in Eq. \eqref{Eq_04} using a Hubbard-Stratonovich transformation
yields the following expression 
\begin{align}\label{Eq_07}
 & \exp\biggl\{ V_{pd}\int d\tau\sum_{i}\sum_{\sigma,\sigma'}\mathcal{A}_{i,\sigma}^{\dagger(3)}\mathcal{A}_{i,\sigma'}^{(3)}\biggr\}\nonumber \\
 & =\int\mathcal{D}[R_{II},\Theta_{II}]\exp\biggl\{\int d\tau\sum_{i,\sigma}\biggl[-\frac{R_{II}^{2}}{2V_{pd}}+R_{II}e^{i\Theta_{II}}\mathcal{A}_{i,\sigma}^{\dagger(3)}\nonumber \\
 & +R_{II}e^{-i\Theta_{II}}\mathcal{A}_{i,\sigma}^{(3)}\biggr]\biggr\},
\end{align}
where $R_{II}e^{i\Theta_{II}}=V_{pd}\sum_{\sigma}\langle\mathcal{A}_{i,\sigma}^{(3)}\rangle$
is a complex order parameter. The mean-field value of the phase $\Theta_{II}$
was determined in Ref. \cite{Varma} as being equal to $\pm\pi/2$.
In what follows, we will choose for simplicity the positive value
of $\Theta_{II}$, since it has been shown in Ref. \cite{Varma} that
this choice minimizes the energy for the present case.

At this point, we would like to point out that we will consider only
the lowest energy band of the noninteracting Hamiltonian defined in
Eq. \eqref{Eq_01}. For physically motivated choices of the parameters
in the present model, the low-energy band will naturally give rise
to a Fermi surface shown in Fig. \ref{Three_Band_Model_03}. The most
singular contribution in this effective model will arise from the
points at the Fermi surface (the so-called hot spots) that represents
the intersection of this surface with the antiferromagnetic zone boundary.
Therefore, we will restrict the analysis of the present model to the
vicinity of these important hot spot points in the considerations
that follow. With this in mind and to set up our notation, we now
define the following $16$-component fermionic spinors 

\vspace{-5cm}

\begin{equation}\label{Eq_08}
d=\begin{pmatrix}\begin{pmatrix}\begin{pmatrix}d_{1}\\
d_{2}
\end{pmatrix}_{\Sigma}\\
\begin{pmatrix}d_{3}\\
d_{4}
\end{pmatrix}_{\Sigma}
\end{pmatrix}_{\Lambda}\\
\begin{pmatrix}\begin{pmatrix}d_{5}\\
d_{6}
\end{pmatrix}_{\Sigma}\\
\begin{pmatrix}d_{7}\\
d_{8}
\end{pmatrix}_{\Sigma}
\end{pmatrix}_{\Lambda}
\end{pmatrix}_{L},p_{x(y)}=\begin{pmatrix}\begin{pmatrix}\begin{pmatrix}p_{x(y)1}\\
p_{x(y)2}
\end{pmatrix}_{\Sigma}\\
\begin{pmatrix}p_{x(y)3}\\
p_{x(y)4}
\end{pmatrix}_{\Sigma}
\end{pmatrix}_{\Lambda}\\
\begin{pmatrix}\begin{pmatrix}p_{x(y)5}\\
p_{x(y)6}
\end{pmatrix}_{\Sigma}\\
\begin{pmatrix}p_{x(y)7}\\
p_{x(y)8}
\end{pmatrix}_{\Sigma}
\end{pmatrix}_{\Lambda}
\end{pmatrix}_{L},
\end{equation}
where $d_{i}=\begin{pmatrix}d_{i,\uparrow}\\
d_{i,\downarrow}
\end{pmatrix}_{\sigma}$ and $p_{x(y)i}=\begin{pmatrix}p_{x(y)i,\uparrow}\\
p_{x(y)i,\downarrow}
\end{pmatrix}_{\sigma}$ are also spinors in the spin space $\sigma$. The symbols $\Sigma$,
$\Lambda$, and $L$ represent independent pseudospin spaces that
are generated by the Pauli matrices \cite{Efetov}. By making use
of the results in Eq. \eqref{Eq_01}, \eqref{Eq_03}, \eqref{Eq_07},
and \eqref{Eq_08} and linearizing the excitation spectrum of three-band
model around the hot spots, we obtain that the total action of the
system yields \begin{widetext} 
\begin{align}\label{Eq_09}
 & \mathcal{S}[p_{x},p_{y},d,\vec{\phi};n_{p},R_{II}]=\mathcal{S}_{0}[p_{x},p_{y},d]+\mathcal{S}_{int}^{(1)}[d,\vec{\phi}]+\mathcal{S}_{int}^{(2)}[p_{x},p_{y},d;n_{p},R_{II}]\nonumber \\
 & =\int\begin{pmatrix}p_{x}^{\dagger}(X), & p_{y}^{\dagger}(X), & d^{\dagger}(X)\end{pmatrix}\begin{pmatrix}\partial_{\tau}+\xi_{p} & \hat{\Gamma}_{1}+\hat{\Gamma}_{2}(-i\nabla) & \hat{\Gamma}_{1x}-\hat{\Gamma}_{2x}i\partial_{x}\\
\hat{\Gamma}_{1}+\hat{\Gamma}_{2}(-i\nabla) & \partial_{\tau}+\xi_{p} & \Gamma_{1y}-\Gamma_{2y}i\partial_{y}\\
\hat{\Gamma}_{1x}^{\dagger}-\hat{\Gamma}_{2x}^{\dagger}i\partial_{x} & \hat{\Gamma}_{1y}^{\dagger}-\hat{\Gamma}_{2y}^{\dagger}i\partial_{y} & \partial_{\tau}+\xi_{d}
\end{pmatrix}\begin{pmatrix}p_{x}(X)\\
p_{y}(X)\\
d(X)
\end{pmatrix}dX\nonumber \\
 & +\frac{1}{2}\int\left[\frac{1}{v_{s}^{2}}(\partial_{\tau}\vec{\phi})^{2}+(\nabla\vec{\phi})^{2}+m_{a}\vec{\phi}^{2}+\frac{g}{2}(\vec{\phi}^{2})^{2}\right]dX+\lambda\int\left[d^{\dagger}(X)\Sigma_{1}\vec{\phi}(X)\vec{\sigma}d(X)\right]dX\nonumber \\
 & +\int\left(\frac{R_{II}^{2}}{V_{pd}}-\frac{n_{p}^{2}}{8}U_{p}\right)dX,
\end{align}
\end{widetext} where $\xi_{p}\equiv\varepsilon_{p}+\frac{n_{p}}{4}U_{p}-\mu$, $\xi_{d}\equiv\varepsilon_{d}-\mu$, and both time and space coordinates have been collected in terms of the variable $X=(\tau,\textbf{r})$. The matrices $\hat{\Gamma}_{1}$, $\hat{\Gamma}_{2}$, $\hat{\Gamma}_{1x(y)}$,
and $\hat{\Gamma}_{2x(y)}$ appearing in Eq. \eqref{Eq_09} are diagonal in the $\Sigma\otimes\Lambda\otimes L$
pseudospin space and depend on all the parameters of the three-band model and also on the order parameter $R_{II}$ for the $\Theta_{II}$-loop current phase (see the Appendix A to check their definition).
Here we follow Ref. \cite{Efetov} and introduce the $32$-component
fermionic spinors in the particle-hole space $\tau$ as 
\begin{align}
 & \Psi=\frac{1}{\sqrt{2}}\begin{pmatrix}d^{*}\\
i\sigma_{2}d
\end{pmatrix}_{\tau},\Psi^{\dagger}=\frac{1}{\sqrt{2}}\begin{pmatrix}-d^{t}, & -d^{\dagger}i\sigma_{2}\end{pmatrix}_{\tau},\label{Eq_10}
\end{align}
\begin{align}
 & P_{x}=\frac{1}{\sqrt{2}}\begin{pmatrix}p_{x}^{*}\\
i\sigma_{2}p_{x}
\end{pmatrix}_{\tau},P_{x}^{\dagger}=\frac{1}{\sqrt{2}}\begin{pmatrix}-p_{x}^{t}, & -p_{x}^{\dagger}i\sigma_{2}\end{pmatrix}_{\tau},\label{Eq_11}\\
 & P_{y}=\frac{1}{\sqrt{2}}\begin{pmatrix}p_{y}^{*}\\
i\sigma_{2}p_{y}
\end{pmatrix}_{\tau},P_{y}^{\dagger}=\frac{1}{\sqrt{2}}\begin{pmatrix}-p_{y}^{t}, & -p_{y}^{\dagger}i\sigma_{2}\end{pmatrix}_{\tau}.\label{Eq_12}
\end{align}
In addition to the fermionic fields defined above, we also introduce
the charge-conjugated vectors as 
\begin{align}\label{Eq_13}
\xoverline[0.5]{\Psi}=\Psi^{\dagger}\tau_{3},\hspace{0.5cm}\xoverline[0.5]{P}_{x}=P_{x}^{\dagger}\tau_{3},\hspace{0.5cm}\xoverline[0.5]{P}_{y}=P_{y}^{\dagger}\tau_{3},
\end{align}
where $\tau_{3}$ is the usual Pauli matrix defined in the $\tau$
space. Hence by making use of these last definitions, the action in
Eq. \eqref{Eq_09} can be naturally rewritten as \begin{widetext}
\begin{align}\label{Eq_14}
 & \mathcal{S}[P_{x},P_{y},\Psi,\vec{\phi};n_{p},R_{II}]\nonumber \\
 & =\int\begin{pmatrix}\xoverline[0.5]{P}_{x}(X), & \xoverline[0.5]{P}_{y}(X), & \xoverline[0.5]{\Psi}(X)\end{pmatrix}\begin{pmatrix}-\partial_{\tau}+\xi_{p}\tau_{3} & \hat{\Gamma}_{1}\tau_{3}-\hat{\Gamma}_{2}(-i\nabla) & \hat{\Gamma}_{1x}\tau_{3}+\hat{\Gamma}_{2x}i\partial_{x}\\
\hat{\Gamma}_{1}\tau_{3}-\hat{\Gamma}_{2}(-i\nabla) & -\partial_{\tau}+\xi_{p}\tau_{3} & \hat{\Gamma}_{1y}\tau_{3}+\hat{\Gamma}_{2y}i\partial_{y}\\
\hat{\Gamma}_{1x}^{\dagger}\tau_{3}+\hat{\Gamma}_{2x}^{\dagger}i\partial_{x} & \hat{\Gamma}_{1y}^{\dagger}\tau_{3}+\hat{\Gamma}_{2y}^{\dagger}i\partial_{y} & -\partial_{\tau}+\xi_{d}\tau_{3}
\end{pmatrix}\begin{pmatrix}P_{x}(X)\\
P_{y}(X)\\
\Psi(X)
\end{pmatrix}dX\nonumber \\
 & +\lambda\int\left[\xoverline[0.5]{\Psi}(X)\Sigma_{1}\vec{\phi}(X)\vec{\sigma}^{t}\Psi(X)\right]dX+\frac{1}{2}\int\left[\frac{1}{v_{s}^{2}}(\partial_{\tau}\vec{\phi})^{2}+(\nabla\vec{\phi})^{2}+m_{a}\vec{\phi}^{2}+\frac{g}{2}(\vec{\phi}^{2})^{2}\right]dX\nonumber \\
 & +\int\left(\frac{R_{II}^{2}}{V_{pd}}-\frac{n_{p}^{2}}{8}U_{p}\right)dX.
\end{align}
\end{widetext}

In order to derive the thermodynamical properties of the present model,
we should first integrate out the bosonic field in the functional
integral 
\begin{align}\label{Eq_15}
\mathcal{Z} & =\int\exp\Bigl\{-\mathcal{S}[P_{x},P_{y},\Psi,\vec{\phi};n_{p},R_{II}]\Bigr\}\mathcal{D}[P_{x},P_{y},\Psi,\vec{\phi}]\nonumber \\
 & =\int\exp\Bigl\{-\mathcal{S}[P_{x},P_{y},\Psi;n_{p},R_{II}]\Bigr\}\mathcal{D}[P_{x},P_{y},\Psi].
\end{align}
However, before proceeding with that, we will neglect from now on
the spin-density-wave interaction $g$ in the present model, since
this coupling effectively renormalizes to zero under the RG flow in
the low-energy limit \cite{Efetov}. As a result, the partition function
of the three-band model may be computed in closed form giving rise
to the low-energy effective action \begin{widetext} 
\begin{align}\label{Eq_16}
 & \mathcal{S}[P_{x},P_{y},\Psi;n_{p},R_{II}]\nonumber \\
 & =\int\begin{pmatrix}\xoverline[0.5]{P}_{x}(X), & \xoverline[0.5]{P}_{y}(X), & \xoverline[0.5]{\Psi}(X)\end{pmatrix}\begin{pmatrix}-\partial_{\tau}+\xi_{p}\tau_{3} & \hat{\Gamma}_{1}\tau_{3}-\hat{\Gamma}_{2}(-i\nabla) & \hat{\Gamma}_{1x}\tau_{3}+\hat{\Gamma}_{2x}i\partial_{x}\\
\hat{\Gamma}_{1}\tau_{3}-\hat{\Gamma}_{2}(-i\nabla) & -\partial_{\tau}+\xi_{p}\tau_{3} & \hat{\Gamma}_{1y}\tau_{3}+\hat{\Gamma}_{2y}i\partial_{y}\\
\hat{\Gamma}_{1x}^{\dagger}\tau_{3}+\hat{\Gamma}_{2x}^{\dagger}i\partial_{x} & \hat{\Gamma}_{1y}^{\dagger}\tau_{3}+\hat{\Gamma}_{2y}^{\dagger}i\partial_{y} & -\partial_{\tau}+\xi_{d}\tau_{3}
\end{pmatrix}\begin{pmatrix}P_{x}(X)\\
P_{y}(X)\\
\Psi(X)
\end{pmatrix}dX\nonumber \\
 & -\frac{\lambda^{2}}{2}\int\left[\xoverline[0.5]{\Psi}(X)\Sigma_{1}\vec{\sigma}^{t}\Psi(X)\right]D(X-X')\left[\xoverline[0.5]{\Psi}(X')\Sigma_{1}\vec{\sigma}^{t}\Psi(X')\right]dXdX'+\int\left(\frac{R_{II}^{2}}{V_{pd}}-\frac{n_{p}^{2}}{8}U_{p}\right)dX.
\end{align}
\end{widetext} 
Here the function $D(X-X')$ that appears as a potential function in the fermionic quartic interaction is the bare bosonic
propagator. Its Fourier transform is given by $D(\omega,\mathbf{k})=(\omega^{2}/v_{s}^{2}+|\mathbf{k}|^{2}+m_{a})^{-1}$
with $m_a$ standing for the spin-wave boson mass that vanishes at the QCP, $v_s$ is the spin-wave velocity and $\omega$ denotes the Matsubara bosonic frequency.

Next, we decouple the fermionic quartic term of the action in Eq. \eqref{Eq_16} by using a composite order parameter $M(X,X')$ for both the quadrupole density wave (QDW) and the $d$-wave singlet superconducting (SSC) orders, as was described in full detail in Ref. \cite{Efetov}. This is achieved by considering the renormalization of bosonic propagator
$D(\omega,\mathbf{k})$ by the fermions at the hot spots which leads to the appearance of the effective spin-wave propagator $D_{eff}(\omega,\mathbf{k})=(\gamma|\omega|+|\mathbf{k}|^{2}+m_{a})^{-1}$,
where $\gamma$ is naturally the Landau damping term. As a consequence, the low-energy effective action that describes the present system may be represented as follows 
\begin{widetext} 
\begin{align}\label{Eq_17}
 & \mathcal{S}_{eff}[P_{x},P_{y},\Psi;n_{p},R_{II},M]\nonumber \\
 & =\int\begin{pmatrix}\xoverline[0.5]{P}_{x}(X), & \xoverline[0.5]{P}_{y}(X), & \xoverline[0.5]{\Psi}(X)\end{pmatrix}\begin{pmatrix}-\partial_{\tau}+\xi_{p}\tau_{3} & \hat{\Gamma}_{1}\tau_{3}-\hat{\Gamma}_{2}(-i\nabla) & \hat{\Gamma}_{1x}\tau_{3}+\hat{\Gamma}_{2x}i\partial_{x}\\
\hat{\Gamma}_{1}\tau_{3}-\hat{\Gamma}_{2}(-i\nabla) & -\partial_{\tau}+\xi_{p}\tau_{3} & \hat{\Gamma}_{1y}\tau_{3}+\hat{\Gamma}_{2y}i\partial_{y}\\
\hat{\Gamma}_{1x}^{\dagger}\tau_{3}+\hat{\Gamma}_{2x}^{\dagger}i\partial_{x} & \hat{\Gamma}_{1y}^{\dagger}\tau_{3}+\hat{\Gamma}_{2y}^{\dagger}i\partial_{y} & -\partial_{\tau}+\xi_{d}\tau_{3}
\end{pmatrix}\begin{pmatrix}P_{x}(X)\\
P_{y}(X)\\
\Psi(X)
\end{pmatrix}dX\nonumber \\
 & -i\int\xoverline[0.5]{\Psi}(X)M(X,X')\Psi(X')dXdX'+\frac{1}{2}\int J^{-1}(X-X')\text{Tr}[M(X,X')\Sigma_{1}M(X',X)\Sigma_{1}]dXdX'\nonumber \\
 & +\int\left(\frac{R_{II}^{2}}{V_{pd}}-\frac{n_{p}^{2}}{8}U_{p}\right)dX,
\end{align}
\end{widetext} where we have written $J(X-X')=3\lambda^{2}D_{eff}(X-X')$ instead of the spin-wave propagator in order to simplify the notation. The order parameter $M(X,X')$ for the QDW/SSC composite order is given by 
\begin{align}
 & M(X,X')=b(X,X')\Sigma_{3}\begin{pmatrix}0 & \hat{u}_{\tau}\\
-\hat{u}_{\tau}^{\dagger} & 0
\end{pmatrix}_{\Lambda},\label{Eq_18}\\
 & \text{with}\hspace{0.5cm}\hat{u}_{\tau}=\begin{pmatrix}\Delta_{-} & \Delta_{+}\\
-\Delta_{+}^{*} & \Delta_{-}^{*}
\end{pmatrix}_{\tau}.\label{Eq_19}
\end{align}
Here $\Delta_{+}$ and $\Delta_{-}$ are, respectively, the $d$-wave
singlet superconducting (SSC) and quadrupole density wave (QDW) components
of the order parameter defined above. We also point out that the matrices
$\hat{u}_{\tau}$ belong to the $SU(2)$ group \cite{Efetov} which
lead to the constraint $|\Delta_{+}|^{2}+|\Delta_{-}|^{2}=1$ involving
both the SSC and QDW sectors. Although we have constructed an effective spin-fermion model
for the CuO$_{2}$ unit cell by considering only the Cu atoms, we point out that the 
QDW/SSC order parameter in Eq. \eqref{Eq_18} does not lead to a charge 
modulation located on the Cu orbitals. In fact, it can be shown \cite{Efetov,Efetov_2,Efetov_3} that this composite
order parameter generates a charge modulation with a checkerboard pattern residing on the oxygen O
sites, which is described by incommensurate wavevectors with respect to the lattice.

The effective action in Eq. \eqref{Eq_17} now has a quadratic form
and the free energy of the system can be obtained as follows: First
one has to integrate out the fermionic fields in the functional integral
for the partition function and then apply the formulae $\text{Tr}\ln G^{-1}=\ln\det(G^{-1})$.
Following this procedure, we determine that the free energy in space-time
coordinates evaluates to 
\begin{align}\label{Eq_20}
 & \frac{F[T,n_{p},R_{II},M]}{T}=-\int\text{Tr}\ln[G^{-1}(X,X')]dXdX'\nonumber \\
 & +\frac{1}{2}\int J^{-1}(X-X')\text{Tr}[M(X,X')\Sigma_{1}M(X',X)\Sigma_{1}]dXdX'\nonumber \\
 & +\int\left(\frac{R_{II}^{2}}{V_{pd}}-\frac{n_{p}^{2}}{8}U_{p}\right)dX,
\end{align}
where the matrix $G^{-1}(X,X')$ is the Fourier transform of $G^{-1}(i\varepsilon_{n},\mathbf{k})$.
This latter function is given by 
\begin{equation}\label{Eq_21}
\begin{pmatrix}-i\varepsilon_{n}+\xi_{p}\tau_{3} & \hat{\Gamma}_{1}\tau_{3}-\hat{\Gamma}_{2}(\mathbf{k}) & \hat{\Gamma}_{1x}\tau_{3}-\hat{\Gamma}_{2x}k_{x}\\
\hat{\Gamma}_{1}^{\dagger}\tau_{3}-\hat{\Gamma}_{2}^{\dagger}(\mathbf{k}) & -i\varepsilon_{n}+\xi_{p}\tau_{3} & \hat{\Gamma}_{1y}\tau_{3}-\hat{\Gamma}_{2y}k_{y}\\
\hat{\Gamma}_{1x}^{\dagger}\tau_{3}-\hat{\Gamma}_{2x}^{\dagger}k_{x} & \hat{\Gamma}_{1y}^{\dagger}\tau_{3}-\hat{\Gamma}_{2y}^{\dagger}k_{y} & -i\varepsilon_{n}+\xi_{d}\tau_{3}-iM(\varepsilon_{n},\mathbf{k})
\end{pmatrix}.
\end{equation}

The self-consistency equation for $b(X,X')$ is derived by minimizing
the free energy $F[T,n_{p},R_{II},M]$ with respect to this order
parameter. As a consequence, we obtain the following equation 
\begin{align}\label{Eq_22}
 & -\text{Tr}\biggl\{\frac{1}{G^{-1}(X,X')}\frac{\partial G^{-1}(X,X')}{\partial b(X,X')}\biggr\}+J^{-1}(X-X')b(X,X')\nonumber \\
 & \times\text{Tr}\biggl\{\Sigma_{3}\begin{pmatrix}0 & \hat{u}_{\tau}\\
-\hat{u}_{\tau}^{\dagger} & 0
\end{pmatrix}_{\Lambda}\Sigma_{1}\Sigma_{3}\begin{pmatrix}0 & \hat{u}_{\tau}\\
-\hat{u}_{\tau}^{\dagger} & 0
\end{pmatrix}_{\Lambda}\Sigma_{1}\biggr\}=0.
\end{align}
By performing the trace operation over the space $\Sigma\otimes\Lambda\otimes L\otimes\tau$
for the second term on the left-hand-side of the equation above, the
order parameter $b(X,X')$ can be simply expressed as 
\begin{align}\label{Eq_23}
 & b(X,X')\nonumber \\
 & =\frac{1}{16}J(X-X')\text{Tr}\biggl\{ G(X,X')\frac{\partial G^{-1}(X,X')}{\partial b(X,X')}\biggr\}\nonumber \\
 & =\frac{1}{16}J(X-X')\text{Tr}\biggl\{ G(X,X')i\Pi_{3}\Sigma_{3}\begin{pmatrix}0 & \hat{u}_{\tau}\\
-\hat{u}_{\tau}^{\dagger} & 0
\end{pmatrix}_{\Lambda}\biggr\},
\end{align}
where $\Pi_{3}$ is a projector for the three-band-model space which
is defined as 
\begin{equation}\label{Eq_24}
\Pi_{3}=\begin{pmatrix}0 & 0 & 0\\
0 & 0 & 0\\
0 & 0 & 1
\end{pmatrix}.
\end{equation}

\noindent At this point, we will make use of the ansatz $b(X,X')=b(X-X')$
{[}and $G(X,X')=G(X-X')${]} and Fourier transform Eq. \eqref{Eq_23}
to momentum-frequency space. As a result, we get the expression 
\begin{align}\label{Eq_25}
b(\varepsilon_{n},\mathbf{k}) & =\frac{T}{16}\sum_{\varepsilon'_{n}}\int J(\varepsilon_{n}-\varepsilon'_{n},\mathbf{k}-\mathbf{k}')\text{Tr}\biggl\{[G^{-1}(i\varepsilon'_{n},\mathbf{k}')]^{-1}\nonumber \\
 & \times\frac{\partial G^{-1}(i\varepsilon'_{n},\mathbf{k}')}{\partial b(\varepsilon'_{n},\mathbf{k}')}\biggr\}\frac{d\mathbf{k}}{(2\pi)^{2}}.
\end{align}
In order to express $b(\varepsilon_{n},\mathbf{k})$ in a convenient
form, we need to evaluate the trace that appears in the above equation.
This problem can be circumvented by using the following identity 
\begin{align}\label{Eq_26}
 & \text{Tr}\biggl\{[G^{-1}(i\varepsilon'_{n},\mathbf{k}')]^{-1}\frac{\partial G^{-1}(i\varepsilon'_{n},\mathbf{k}')}{\partial b(\varepsilon'_{n},\mathbf{k}')}\biggr\}\nonumber \\
 & =\frac{1}{\det[G^{-1}(i\varepsilon'_{n},\mathbf{k}')]}\frac{\partial\det[G^{-1}(i\varepsilon'_{n},\mathbf{k}')]}{\partial b(\varepsilon'_{n},\mathbf{k}')}.
\end{align}
Then, by substituting Eq. \eqref{Eq_26} into Eq. \eqref{Eq_25},
we finally arrive at the self-consistency equation 
\begin{align}\label{Eq_27}
b(\varepsilon_{n},\mathbf{k}) & =\frac{3\lambda^{2}T}{16}\sum_{\varepsilon'_{n}}\int\frac{D_{eff}(\varepsilon_{n}-\varepsilon'_{n},\mathbf{k}-\mathbf{k}')}{\det[G^{-1}(i\varepsilon'_{n},\mathbf{k}')]}\nonumber \\
 & \times\frac{\partial\det[G^{-1}(i\varepsilon'_{n},\mathbf{k}')]}{\partial b(\varepsilon'_{n},\mathbf{k}')}\frac{d\mathbf{k}}{(2\pi)^{2}},
\end{align}
where we have set $J(\varepsilon_{n}-\varepsilon'_{n},\mathbf{k}-\mathbf{k}')=3\lambda^{2}D_{eff}(\varepsilon_{n}-\varepsilon'_{n},\mathbf{k}-\mathbf{k}')$.

We now turn our attention to the evaluation of $\det[G^{-1}(i\varepsilon_{n},\mathbf{k})]$.
In order to do this, we will need to use the set of determinant formulas
\begin{align}
 & \det\begin{pmatrix}\hat{A} & \hat{B}\\
\hat{C} & \hat{D}
\end{pmatrix}=\det(\hat{A})\det(\hat{D}-\hat{C}\hat{A}^{-1}\hat{B}),\label{Eq_28}\\
 & \det(\hat{A}\otimes\hat{D})=[\det(\hat{A})]^{m}[\det(\hat{D})]^{n},\label{Eq_29}
\end{align}
where $\hat{A}$ and $\hat{D}$ are, respectively, $n$- and $m$-square
matrices and $\det(\hat{A})$ is different from zero. In this way,
by neglecting the SSC sector of the QDW/SSC order parameter ($\Delta_{+}=0$)
and applying these reduction formulas to the matrix $G^{-1}(i\varepsilon_{n},\mathbf{k})$,
we obtain, after some algebraic manipulations, that $\det[G^{-1}(i\varepsilon_{n},\mathbf{k})]$
evaluates formally to 
\begin{equation}\label{Eq_30}
\det[G^{-1}(i\varepsilon_{n},\mathbf{k})]=\prod\limits _{l=1}^{2}\prod\limits _{m=1}^{2}\mathcal{D}_{l}^{(m)}(i\varepsilon_{n},\mathbf{k}),
\end{equation}
where $\mathcal{D}_{l}^{(m)}(i\varepsilon_{n},\mathbf{k})$ are well-behaved
functions of the three-band-model parameters, which are computed in
detailed form in Appendices B and C. Thus, by inserting the result
displayed in Eq. \eqref{Eq_30} into Eq. \eqref{Eq_27}, the mean-field
equation for $b(\varepsilon_{n},\mathbf{k})$ in terms of $\mathcal{D}_{l}^{(m)}(i\varepsilon_{n},\mathbf{k})$
finally reads 
\begin{align}\label{Eq_31}
b(\varepsilon_{n},\mathbf{k}) & =\frac{3\lambda^{2}T}{16}\sum\limits _{l,m=1}^{2}\sum\limits _{\varepsilon'_{n}}^ {}\int\dfrac{D_{eff}(\varepsilon_{n}-\varepsilon'_{n},\mathbf{k}-\mathbf{k}')}{\mathcal{D}_{l}^{(m)}(i\varepsilon'_{n},\mathbf{k}')}\nonumber \\
 & \times\dfrac{\partial\mathcal{D}_{l}^{(m)}(i\varepsilon'_{n},\mathbf{k}')}{\partial b(\varepsilon'_{n},\mathbf{k}')}\frac{d\mathbf{k}'}{(2\pi)^{2}}.
\end{align}
We note that, since we have set $\Delta_{+}=0$, the above self-consistency
equation describes only the QDW sector of the fluctuations in the
present system.

\begin{figure*}[t]
\begin{center}
\includegraphics[width=1.0 \linewidth]{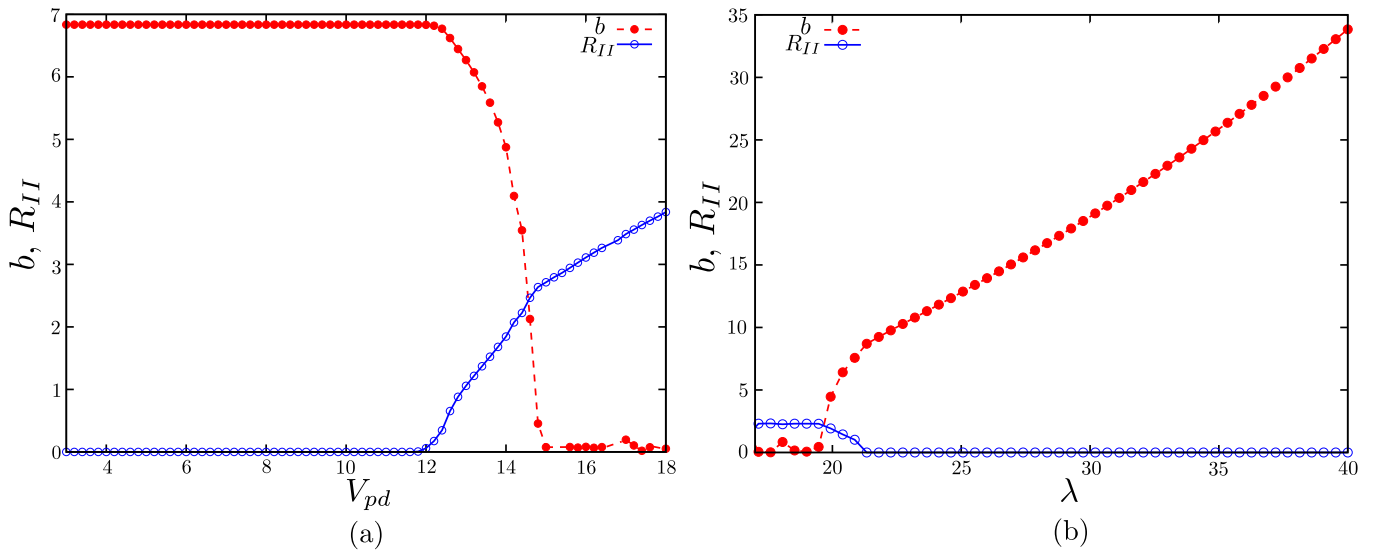}
\caption{(Color online) (a) Mean-field values of $R_{II}$ and $b$ as a function
of the nearest-neighbor interaction $V_{pd}$ in the limit of zero
temperature for $\lambda=20$. (b) Mean-field values of $R_{II}$
and $b$ as a function of the spin-fermion coupling $\lambda$ in
the limit of zero temperature for $V_{pd}=14$. Both solutions in
(a) and (b) were obtained by performing numerical integration in momentum
space of the self-consistency equations given by Eqs. \eqref{Eq_31}
and \eqref{Eq_33} with a mesh of $320\times320$ points in the Brillouin
zone. Here $m_{a}=10^{-2}$, $\gamma=10^{-5}$ and the other interactions
are set to $t_{pd}=1$, $t_{pp}=0.5$, $U_{p}=3$, and $\varepsilon_{d}-\varepsilon_{p}=3$.
The fermionic density on the O orbital is given by $n_{p}=0.6$ and
the position of the hot spots is such that $\delta=0.93$.}\label{Mean_Field_III}
\end{center} 
\end{figure*}

As a consequence of the result in Eq. \eqref{Eq_30}, we determine
after Fourier transforming the right-hand-side of Eq. \eqref{Eq_20}
that the free energy of the present model has the following analytical
form 
\begin{align}\label{Eq_32}
 & F[T,n_{p},R_{II},b]\nonumber \\
 & =-T\sum\limits _{l,m=1}^{2}\sum_{\varepsilon_{n}}\int\ln\bigl[\mathcal{D}_{l}^{(m)}(i\varepsilon_{n},\mathbf{k})\bigr]\frac{d\mathbf{k}}{(2\pi)^{2}}\nonumber \\
 & +\frac{8T}{3\lambda^{2}}\sum_{\varepsilon_{n}}\int b(\varepsilon_{n},\mathbf{k})\frac{d\mathbf{k}}{(2\pi)^{2}}\biggl[T\sum_{\varepsilon'_{n}}\int b(\varepsilon'_{n},\mathbf{k}')\nonumber \\
 & \times D_{eff}^{-1}(\varepsilon_{n}-\varepsilon'_{n},\mathbf{k}-\mathbf{k}')\frac{d\mathbf{k}'}{(2\pi)^{2}}\biggr]+\frac{R_{II}^{2}}{V_{pd}}-\frac{n_{p}^{2}}{8}U_{p},
\end{align}
where we have set the volume of the system to unity. In order to self-consistently
determine the mean-field order parameter $R_{II}$, we need also minimize
the free energy with respect to it. In this way, the self-consistency
equation for $R_{II}$ in turn reads 
\begin{equation}\label{Eq_33}
R_{II}=\frac{V_{pd}T}{2}\sum\limits _{l,m=1}^{2}\sum_{\varepsilon_{n}}\int\frac{1}{\mathcal{D}_{l}^{(m)}(i\varepsilon_{n},\mathbf{k})}\frac{\partial\mathcal{D}_{l}^{(m)}(i\varepsilon_{n},\mathbf{k})}{\partial R_{II}}\frac{d\mathbf{k}}{(2\pi)^{2}}.
\end{equation}
The solutions of both Eqs. \eqref{Eq_31} and \eqref{Eq_33} will
be obtained in the next section, following a numerical procedure described
in great detail in Appendices B and C.

\section{Mean-field results}

In order to investigate the interplay between both $\Theta_{II}$-loop-current (LC)
and QDW orders in the present three-band model, we solve numerically
the mean-field equations for $R_{II}$ and $b$. The present numerical
approach consists in the discretization of the Brillouin zone with
a mesh of $320\times320$ points. We also make the assumption that
the order parameter $b(\varepsilon_{n},\mathbf{k})$ does not depend
crucially on the frequency and momentum. In this way, we will only
investigate the ground state properties of the present model, which
therefore allows us to evaluate the Matsubara sums
that appear in the mean-field equations exactly. We perform this calculation
by either varying the spin-fermion coupling $\lambda$ or the nearest-neighbor
interaction $V_{pd}$ between O and Cu orbitals. In addition, we fix
all other couplings in the theory. The corresponding results are shown
in Figs. \ref{Mean_Field_III}(a) and (b). For physically realistic
parameters in the present model (here we choose, e.g., $m_{a}=10^{-2}$,
$\gamma=10^{-5}$, $t_{pd}=1$, $t_{pp}=0.5$, $U_{p}=3$, and $\varepsilon_{d}-\varepsilon_{p}=3$),
we observe in Fig. \ref{Mean_Field_III}(a) that the order parameter
$R_{II}$ grows continuously from zero to positive values as the interaction
$V_{pd}$ is increased for $\lambda$ fixed. It can be very interesting at this point to make a rough 
estimate of the magnetic moment associated with the loop currents described by $R_{II}$ obtained here at mean-field level.  
From Fig. \ref{Mean_Field_III}(a), we can estimate numerically that the ratio of the critical parameters is given approximately by $(R_{II}^c/V_{pd}^c)\sim 0.2$. 
Hence, by following the same calculation procedure that was explained in detail in Ref. \cite{Thomale}, we may conclude that the $\Theta_{II}$-loop-current phase 
in our present theory yields a magnetic moment per unit-cell of approximately $M_{LC}\sim 0.19 \mu_B$. Quite surprisingly, this result agrees qualitatively 
with the experimental estimate of $M_{exp}\sim 0.05 \mu_B - 0.1 \mu_B$ found by Fauqué \emph{et al.} \cite{Fauque} using spin-polarized neutron scattering experiments.
In Fig. \ref{Mean_Field_III}(a), it can also be seen 
that the QDW order parameter $b$, by contrast, vanishes as the interaction $V_{pd}$ becomes
stronger. Moreover, one can note in the same figure
a narrow region where both order parameters can be finite for moderate
$V_{pd}$, indicating that the present three-band model could in principle
accommodate a coexisting phase involving both time-reversal (LC order)
and translational symmetry breaking (QDW order), but as can be inferred
from Fig. \ref{Mean_Field_III}(a) this apparently occurs for somewhat
fine-tuned interactions. 

We can also analyze the behavior of the same order parameters as a
function of spin-fermion coupling $\lambda$, when we keep instead
the interaction $V_{pd}$ fixed. The corresponding results are depicted
in Fig. \ref{Mean_Field_III}(b). As a result, we find that the LC
order parameter $R_{II}$ is finite below a threshold of $\lambda$
and then is clearly suppressed when this interaction becomes larger.
Once more, the behavior of the QDW order parameter $b$ is essentially
the opposite one, namely, it grows from zero to finite values as the
spin-fermion interaction becomes stronger. In an analogous way to
the previous case, there is also a very narrow window where both phases
may coexist for moderate $\lambda$ and $V_{pd}$. Despite this, the
generic behavior which can inferred from both figures is that the
LC order appears to be detrimental to the QDW order and vice-versa.
In other words, we may conclude at this point that, for a large majority
of initial choices for the couplings $V_{pd}$ and $\lambda$ within
the present three-band model, there is a strong tendency for the above
two orders not to coexist, at least at mean-field level.

In order to analyze the sensitivity of the above mean-field results to changes in the physical parameters of the three-band model, we have also investigated its properties 
with respect to varying both the spin-wave bosonic mass $m_{a}$ and the orbital-energy transfer 
$\varepsilon_{d}-\varepsilon_{p}$. As the strength of $m_{a}$ becomes larger (which corresponds naturally
to shorter SDW correlation lengths), we obtain a clear tendency for both critical interactions (i.e. $\lambda_c$ and $V_{pd}^c$) to increase even further in our numerical data. 
This result would of course lie beyond the regime of applicability of a mean-field approach to the present model and other complementary methods that include quantum fluctuation effects should
be used to describe such a regime. 
In addition to this, we have also examined the dependence of our results with respect to changes in the orbital-energy transfer of the model.
As a consequence, we were able to establish numerically that, as the difference $\varepsilon_{d}-\varepsilon_{p}$ is reduced towards zero, 
the critical interactions $\lambda_c$ and $V_{pd}^c$ also display a tendency to increase further within the present approach.

\section{Conclusions}

In the present work, we have performed a consistent mean-field calculation
for the three-band (Emery) model relevant to the phenomenology of
the underdoped cuprates. We have shown that a low-energy effective
description of this model may indeed exhibit both the $\Theta_{II}$-loop-current
order first proposed by Varma\cite{Varma} and the so-called QDW which
arises from an emergent $SU(2)$ pseudospin symmetry that exists in
the spin-fermion model \cite{Metlitski,Efetov}. As a result, we have
obtained that the above two order parameters have a tendency to be detrimental to each
other, at least at mean-field level.  

We would like to point out that the mean-field values of the
critical interactions to obtain these two phases are relatively large compared 
with some physical parameters of the three-band model. This is expected to be an artifact of the mean-field approach and, for this reason, 
other complementary methods (such as, e.g., renormalization group techniques that include quantum fluctuation effects)
should be used in order to establish a quantitative agreement between the present model and the experimental 
data. It is also important to mention that there are other works in the literature,
which analyzed three-band models using weak-coupling diagrammatic perturbative calculations \cite{Bulut_2,Thomson}. They have confirmed
that the QDW order with a $d$-wave form factor investigated in the present work turns out to be more stable than the experimentally
observed charge order with a modulation along the axial directions. Here, we have shown that the three-band model 
can also accommodate a $\Theta_{II}$-loop-current phase that breaks time-reversal symmetry, which seemingly acts against the QDW order. This suggests an appealing scenario
where the $\Theta_{II}$-loop-current-order strongly
competes with the QDW order, with one order having a tendency to suppress
the other (and vice-versa) in the present model.
This clearly opens an interesting avenue for future research and may help
rule out recent competing (and mutually exclusive) interpretations
of the universal phenomenon of the pseudogap phase displayed in the
underdoped cuprates, in light of the many highly-precise experiments
performed in those materials in the last years.

\section{Acknowledgments}

One of us (V. S. de C.) would like to thank a fellowship from CAPES
(No. 99999.000324/2014-00) under the program `Science Without Borders'. He also acknowledges the kind hospitality of the IPhT in CEA-Saclay
during his stay and useful conversations with Salviano Le\~{a}o. H. F. acknowledges the support from FAPEG under the
grant No. 201210267001167. This work was supported
partly by LabEx PALM (ANR-10-LABX-0039- PALM), by the ANR project
UNESCOS ANR-14-CE05-0007, as well as the grant Ph743-12 of the COFECUB
which enabled frequent visits to the International Institute of Physics (IIP), in Natal. X.M. and T.K. also
acknowledge the support of CAPES and funding from the IIP. 

\appendix

\section{Definition of the $\hat{\Gamma}_{i}$ matrices}

The matrices $\hat{\Gamma}_{1}$, $\hat{\Gamma}_{2}$, $\hat{\Gamma}_{1x}$,
$\hat{\Gamma}_{2x}$, $\hat{\Gamma}_{1y}$, and $\hat{\Gamma}_{2y}$
that appear throughout this work are defined by linearizing the functions of the three-band 
model around the hot spots depicted in Fig. \ref{Three_Band_Model_03}. The structure of the
resulting matrices can then be simplified by resorting to a representation based on Pauli matrices 
defined in distinct pseudospin spaces \cite{Efetov}, which are denoted by $\Sigma$, $\Lambda$, and $L$. 
Technically speaking, the pseudospin space $\Sigma$ connects hot spots that can be mapped onto each other by 
the antiferromagnetic wavevector $\mathbf{Q}=(\pi,\pi)$. Different pairs of hot spots connected by the wavevector $\mathbf{Q}$ are mapped onto each other by the pseudospin space $\Lambda$. Lastly, the pseudospin space $L$ connects orthogonal quartet of hot spots. Following these definitions, the matrices of the three-band 
model can be simply written as
\begin{align}
\hat{\Gamma}_{1}\  & =-2t_{pp}\cos\delta\ \mathbb{1}_{\Sigma}\otimes\mathbb{1}_{\Lambda}\otimes\mathbb{1}_{L},\label{Eq_A1}\\
\hat{\Gamma}_{2}\  & =t_{pp}(\sin\delta\Lambda_{3}\otimes L_{3}-\Sigma_{3}\otimes\Lambda_{3})i\partial_{x}\nonumber \\
 & -t_{pp}(\sin\delta\Lambda_{3}+\Sigma_{3}\otimes\Lambda_{3}\otimes L_{3})i\partial_{y},\label{Eq_A2}\\
\hat{\Gamma}_{1x} & =\gamma_{1}e^{-i\varphi\Lambda_{3}\otimes L_{3}}+\gamma_{2}e^{i\theta\Lambda_{3}\otimes L_{3}}\Sigma_{3}\otimes L_{3},\label{Eq_A3}\\
\hat{\Gamma}_{2x} & =-\frac{1}{2}\gamma_{1}e^{-i\varphi\Lambda_{3}\otimes L_{3}}\Sigma_{3}\otimes\Lambda_{3}+\frac{1}{2}\gamma_{2}e^{i\theta\Lambda_{3}\otimes L_{3}}\Lambda_{3}\otimes L_{3},\label{Eq_A4}\\
\hat{\Gamma}_{1y} & =\gamma_{1}e^{i\varphi\Lambda_{3}}-\gamma_{2}e^{-i\theta\Lambda_{3}}\Sigma_{3}\otimes L_{3},\label{Eq_A5}\\
\hat{\Gamma}_{2y} & =-\frac{1}{2}\gamma_{1}e^{i\varphi\Lambda_{3}}\Sigma_{3}\otimes\Lambda_{3}\otimes L_{3}+\frac{1}{2}\gamma_{2}e^{-i\theta\Lambda_{3}}\Lambda_{3},\label{Eq_A6}
\end{align}
where $\delta=(K_{+}-K_{-})/2$ and $\mathbb{1}_{\Sigma}$, $\mathbb{1}_{\Lambda}$,
and $\mathbb{1}_{L}$ are, respectively, the identity matrices in
the $\Sigma$, $\Lambda$, and $L$ pseudospin spaces. The parameters
$\varphi$, $\theta$, $\gamma_{1}$, and $\gamma_{2}$ are defined
as 
\begin{align}
 & \tan\varphi=\frac{R_{II}}{2t_{pd}}\tan\biggl(\frac{\delta}{2}\biggr),\label{Eq_A7}\\
 & \tan\theta=\frac{R_{II}}{2t_{pd}}\cot\biggl(\frac{\delta}{2}\biggr),\label{Eq_A8}\\
 & \gamma_{1}=\biggl[2t_{pd}^{2}\cos^{2}\biggl(\frac{\delta}{2}\biggr)+\frac{R_{II}^{2}}{2}\sin^{2}\biggl(\frac{\delta}{2}\biggr)\biggr]^{1/2},\label{Eq_A9}\\
 & \gamma_{2}=\biggl[2t_{pd}^{2}\sin^{2}\biggl(\frac{\delta}{2}\biggr)+\frac{R_{II}^{2}}{2}\cos^{2}\biggl(\frac{\delta}{2}\biggr)\biggr]^{1/2}.\label{Eq_A10}
\end{align}

\section{Evaluation of the Matsubara sums for the mean-field equations}

\subsection{Quadrupole density wave (QDW) order parameter}

In order to compute the Matsubara sum in Eq. \eqref{Eq_31}, we will
consider that the QDW order parameter does not depend on both the
frequency and the momentum. In this manner, we can rewrite this equation
as 
\begin{equation}\label{Eq_B1}
b(T)=\frac{3\lambda^{2}T}{16}\sum\limits _{l,m=1}^{2}\sum\limits _{\varepsilon_{n}}^ {}\int\dfrac{D_{eff}(\varepsilon_{n},\mathbf{k})}{\mathcal{D}_{l}^{(m)}(i\varepsilon_{n},\mathbf{k})}\dfrac{\partial\mathcal{D}_{l}^{(m)}(i\varepsilon_{n},\mathbf{k})}{\partial b}\frac{d\mathbf{k}}{(2\pi)^{2}},
\end{equation}
where we have not written explicitly the full dependence of $b(T)$
to not clutter up the notation.

\begin{figure*}[!htb]
\begin{center}
\includegraphics[width=0.70 \linewidth]{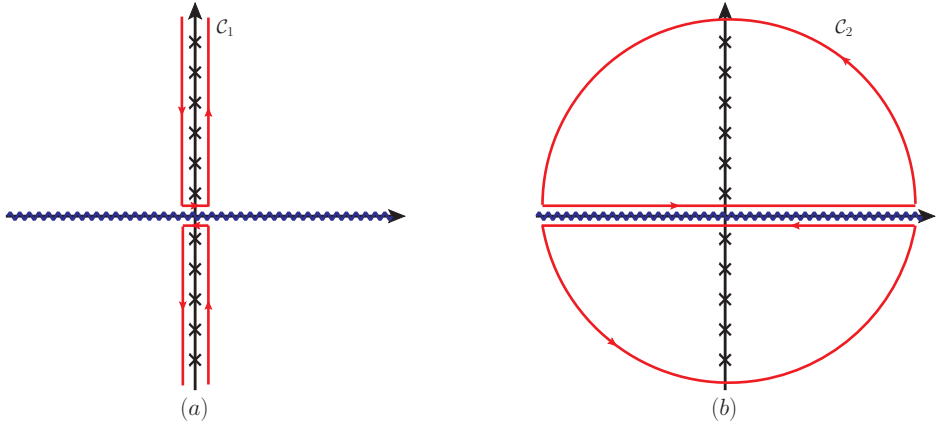}
\caption{($a$) Integration contour $\mathcal{C}_{1}$ for the evaluation of
the Matsubara sum appearing in the mean-field equation for the QDW
order parameter. The crosses ($\times$) represent the poles of the
Fermi-Dirac distribution function $n_{F}(z)$ and the wavy-blue line
at $\operatorname{Im}(z)=0$ is the branch cut of $D_{eff}(-iz,\mathbf{k})$.
($b$) In order to perform the Matsubara sum in this case, the integration
contour $\mathcal{C}_{1}$ can be distorted such that it transforms
into the contour $\mathcal{C}_{2}$ that avoids the poles of $n_{F}(z)$
and also the branch cut.}\label{Three_Band_Model_04}
\end{center}
\end{figure*}

There is a subtlety to obtain the analytic continuation of the effective
bosonic propagator $D_{eff}(\varepsilon_{n},\mathbf{k})$ since this
function depends on $|\omega|$ which is not well-defined for complex
numbers. To circumvent that, we make use of the two integral formulas
\begin{align}
|\omega| & =-\frac{i\omega}{\pi}\int_{-\infty}^{\infty}\frac{dx}{x-i\omega},\label{Eq_B2}\\
\operatorname{sgn}(\omega) & =-\frac{i}{\pi}\int_{-\infty}^{\infty}\frac{dx}{x-i\omega}.\label{Eq_B3}
\end{align}
As a result, the analytic continuation of $D_{eff}(\varepsilon_{n},\mathbf{k})$
becomes 
\begin{equation}\label{Eq_B4}
D_{eff}(-iz,\mathbf{k})=\frac{1}{-i\gamma z\operatorname{sgn}[\operatorname{Im}(z)]+|\mathbf{k}|^{2}+m_{a}}.
\end{equation}
As may be easily concluded, $D_{eff}(-iz,\mathbf{k})$ is not analytic
in the entire complex plane. Indeed, it possesses a branch cut (see
Fig. \ref{Three_Band_Model_04}) which must be avoided when performing
complex integration. As a result, we obtain that Eq. \eqref{Eq_B1}
may be rewritten as 
\begin{align}\label{Eq_B5}
 & b(T)\nonumber \\
 & =\frac{3\lambda^{2}}{16}\sum\limits _{l,m=1}^{2}\int\frac{d\mathbf{k}}{(2\pi)^{2}}\biggl\{-\frac{1}{2\pi i}\oint_{\mathcal{C}_{1}}dzn_{F}(z)D_{eff}(-iz,\mathbf{k})\nonumber \\
 & \times\biggl[\dfrac{1}{h_{l}^{(m)}(z,\mathbf{k})}\dfrac{\partial h_{l}^{(m)}(z,\mathbf{k})}{\partial b}+\dfrac{1}{\xoverline[0.8]{h}_{l}^{(m)}(z,\mathbf{k})}\dfrac{\partial\xoverline[0.8]{h}_{l}^{(m)}(z,\mathbf{k})}{\partial b}\biggr]\biggr\},
\end{align}
where we have used the result $\mathcal{D}_{l}^{(m)}(z,\mathbf{k})=h_{l}^{(m)}(z,\mathbf{k})\xoverline[0.8]{h}_{l}^{(m)}(z,\mathbf{k})$
(see the Appendix C for details). Here $h_{l}^{(m)}(z,\mathbf{k})$
and $\dfrac{\partial h_{l}^{(m)}(z,\mathbf{k})}{\partial b}$ are
both polynomials with respect to $z$ and the relation between their
degrees is the following 
\begin{equation}\label{Eq_B6}
\text{deg}\biggl[\dfrac{\partial h_{l}^{(m)}(z,\mathbf{k})}{\partial b}\biggr]<\text{deg}[h_{l}^{(m)}(z,\mathbf{k})].
\end{equation}
As expected, a similar inequality holds for $\xoverline[0.8]{h}_{l}^{(m)}(z,\mathbf{k})$
and $\dfrac{\partial\xoverline[0.8]{h}_{l}^{(m)}(z,\mathbf{k})}{\partial b}$.
The main consequence of the result in Eq. \eqref{Eq_B6} is that we
can have a series expansion of the form 
\begin{align}\label{Eq_B7}
 & \dfrac{1}{h_{l}^{(m)}(z,\mathbf{k})}\dfrac{\partial h_{l}^{(m)}(z,\mathbf{k})}{\partial b}+\dfrac{1}{\xoverline[0.8]{h}_{l}^{(m)}(z,\mathbf{k})}\dfrac{\partial\xoverline[0.8]{h}_{l}^{(m)}(z,\mathbf{k})}{\partial b}\nonumber \\
 & =\sum\limits _{n=1}^{N}\dfrac{\Delta_{l,n}^{(m)}(\mathbf{k})}{z-\xi_{l,n}^{(m)}(\mathbf{k})},
\end{align}
where $N\equiv\frac{3}{4}\text{dim}(\Sigma\otimes\Lambda\otimes L\otimes\tau)$
is equal to twelve and $\xi_{l,n}^{(m)}(\mathbf{k})$ represent both
the roots of $h_{l}^{(m)}(z,\mathbf{k})$ ($1\leqslant n\leqslant N/2$)
and $\xoverline[0.8]{h}_{l}^{(m)}(z,\mathbf{k})$ ($N/2+1\leqslant n\leqslant N$).
The coefficients $\Delta_{l,n}^{(m)}(\mathbf{k})$ are calculated
as 
\begin{align}
\Delta_{l,n}^{(m)}(\mathbf{k}) & =\dfrac{\dfrac{\partial h_{l}^{(m)}(z,\mathbf{k})}{\partial b}\bigg|_{z=\xi_{l,n}^{(m)}(\mathbf{k})}}{\dfrac{\partial h_{l}^{(m)}(z,\mathbf{k})}{\partial z}\bigg|_{z=\xi_{l,n}^{(m)}(\mathbf{k})}},\hspace{0.5cm}1\leqslant n\leqslant\frac{N}{2};\label{Eq_B8}\\
\Delta_{l,n}^{(m)}(\mathbf{k}) & =\dfrac{\dfrac{\partial\xoverline[0.8]{h}_{l}^{(m)}(z,\mathbf{k})}{\partial b}\bigg|_{z=\xi_{l,n}^{(m)}(\mathbf{k})}}{\dfrac{\partial\xoverline[0.8]{h}_{l}^{(m)}(z,\mathbf{k})}{\partial z}\bigg|_{z=\xi_{l,n}^{(m)}(\mathbf{k})}},\hspace{0.5cm}\frac{N}{2}+1\leqslant n\leqslant N.\label{Eq_B9}
\end{align}

Then by substituting Eq. \eqref{Eq_B7} into Eq. \eqref{Eq_B5}, we
obtain that the mean-field equation for $b(T)$ assumes the form 
\begin{align}\label{Eq_B10}
b(T) & =\frac{3\lambda^{2}}{16}\sum\limits _{l,m=1}^{2}\sum\limits _{n=1}^{N}\int\Delta_{l,n}^{(m)}(\mathbf{k})\biggl[-\frac{1}{2\pi i}\oint_{\mathcal{C}_{1}}dzn_{F}(z)\nonumber \\
 & \times\dfrac{D_{eff}(-iz,\mathbf{k})}{z-\xi_{l,n}^{(m)}(\mathbf{k})}\biggr]\frac{d\mathbf{k}}{(2\pi)^{2}}.
\end{align}
The complex integral between brackets is computed by changing the
integration contour from $\mathcal{C}_{1}$ to $\mathcal{C}_{2}$
(see Fig. \ref{Three_Band_Model_04}), i.e., 
\begin{align}\label{Eq_B11}
 & -\frac{1}{2\pi i}\oint_{\mathcal{C}_{1}}dzn_{F}(z)\dfrac{D_{eff}(-iz,\mathbf{k})}{z-\xi_{l,n}^{(m)}(\mathbf{k})}\nonumber \\
 & =-\frac{1}{2\pi i}\int_{-\infty}^{\infty}dxn_{F}(x)\biggl[\frac{1}{-i\gamma x+|\mathbf{k}|^{2}+m_{a}}\frac{1}{x^{+}-\xi_{l,n}^{(m)}(\mathbf{k})}\nonumber \\
 & -\frac{1}{i\gamma x+|\mathbf{k}|^{2}+m_{a}}\frac{1}{x^{-}-\xi_{l,n}^{(m)}(\mathbf{k})}\biggr],
\end{align}
where $x^{\pm}=x\pm i\eta$ and $\eta\rightarrow0^{+}$. At this point,
we employ the Dirac identity 
\begin{equation}\label{Eq_B12}
\frac{1}{x\pm i\eta}=\mp i\pi\delta(x)+\mathcal{P}\bigg(\frac{1}{x}\biggr),
\end{equation}
with $\mathcal{P}$ standing for the Cauchy principal value in order
to obtain the following 
\begin{align}\label{Eq_B13}
 & -\frac{1}{2\pi i}\oint_{\mathcal{C}_{1}}dzn_{F}(z)\dfrac{D_{eff}(-iz,\mathbf{k})}{z-\xi_{l,n}^{(m)}(\mathbf{k})}\nonumber \\
 & =\frac{|\mathbf{k}|^{2}+m_{a}}{[|\mathbf{k}|^{2}+m_{a}]^{2}+\gamma^{2}[\xi_{l,n}^{(m)}(\mathbf{k})]^{2}}n_{F}[\xi_{l,n}^{(m)}(\mathbf{k})]\nonumber \\
 & -\frac{\gamma}{\pi}\mathcal{P}\int_{-\infty}^{\infty}dx\frac{xn_{F}(x)}{[|\mathbf{k}|^{2}+m_{a}]^{2}+\gamma^{2}x^{2}}\frac{1}{x-\xi_{l,n}^{(m)}(\mathbf{k})}.
\end{align}

Finally after inserting the result in Eq. \eqref{Eq_B13} into Eq.
\eqref{Eq_B10}, the mean-field equation for the QDW order parameter
at finite temperature can be simply expressed as 
\begin{align}\label{Eq_B14}
 b(T) & =\frac{3\lambda^{2}}{16}\sum\limits _{l,m=1}^{2}\sum\limits _{n=1}^{N}\int\biggl\{\frac{|\mathbf{k}|^{2}+m_{a}}{[|\mathbf{k}|^{2}+m_{a}]^{2}+\gamma^{2}[\xi_{l,n}^{(m)}(\mathbf{k})]^{2}}\nonumber \\
 & \times n_{F}[\xi_{l,n}^{(m)}(\mathbf{k})]-\frac{\gamma}{\pi}\mathcal{P}\int_{-\infty}^{\infty}dx\frac{xn_{F}(x)}{[|\mathbf{k}|^{2}+m_{a}]^{2}+\gamma^{2}x^{2}}\nonumber \\
 & \times\frac{1}{x-\xi_{l,n}^{(m)}(\mathbf{k})}\biggr\}\Delta_{l,n}^{(m)}(\mathbf{k})\frac{d\mathbf{k}}{(2\pi)^{2}}.
\end{align}
In the limit of $T\rightarrow0$, the Fermi-Dirac distribution
function $n_{F}(x)$ becomes the step function $\theta(-x)$. As a
result, the integral in Eq. \eqref{Eq_B14} involving the Cauchy principal
value evaluates to 
\begin{align}\label{Eq_B15}
 & \lim_{T\rightarrow0}\mathcal{P}\int_{-\infty}^{\infty}dx\frac{xn_{F}(x)}{[|\mathbf{k}|^{2}+m_{a}]^{2}+\gamma^{2}x^{2}}\frac{1}{x-\xi_{l,n}^{(m)}(\mathbf{k})}\nonumber \\
 & =\frac{\pi}{2\gamma}\frac{|\mathbf{k}|^{2}+m_{a}}{(|\mathbf{k}|^{2}+m_{a})^{2}+\gamma^{2}[\xi_{l,n}^{(m)}(\mathbf{k})]^{2}}\nonumber \\
 & +\frac{\xi_{l,n}^{(m)}(\mathbf{k})}{(|\mathbf{k}|^{2}+m_{a})^{2}+\gamma^{2}[\xi_{l,n}^{(m)}(\mathbf{k})]^{2}}\ln\biggl[\frac{\gamma|\xi_{l,n}^{(m)}(\mathbf{k})|}{|\mathbf{k}|^{2}+m_{a}}\biggr],
\end{align}
where now $m_{a}=m_{a}(T=0)$ is the zero-temperature bosonic mass.
Hence in this limit the mean-field equation for the QDW order parameter
is given by 
\begin{align}\label{Eq_B16}
b(T=0) & =-\frac{3\lambda^{2}}{32}\sum\limits _{l,m=1}^{2}\sum\limits _{n=1}^{N}\int\biggl\{\frac{|\mathbf{k}|^{2}+m_{a}}{(|\mathbf{k}|^{2}+m_{a})^{2}+\gamma^{2}[\xi_{l,n}^{(m)}(\mathbf{k})]^{2}}\nonumber \\
 & \times\operatorname{sgn}[\xi_{l,n}^{(m)}(\mathbf{k})]+\frac{2}{\pi}\frac{\gamma\xi_{l,n}^{(m)}(\mathbf{k})}{(|\mathbf{k}|^{2}+m_{a})^{2}+\gamma^{2}[\xi_{l,n}^{(m)}(\mathbf{k})]^{2}}\nonumber \\
 & \times\ln\biggl[\frac{\gamma|\xi_{l,n}^{(m)}(\mathbf{k})|}{|\mathbf{k}|^{2}+m_{a}}\biggr]\biggr\}\Delta_{l,n}^{(m)}(\mathbf{k})\frac{d\mathbf{k}}{(2\pi)^{2}},
\end{align}
where we have used the identity $\theta(-x)=\frac{1}{2}[1-\operatorname{sgn}(x)]$
in order to simplify the above equation.

\subsection{$\Theta_{II}$-loop-current (LC) order parameter}

The mean-field equation for the loop-current order parameter can be
simplified following the same procedure outlined above. First of all,
we transform the Matsubara sum in Eq. \eqref{Eq_33} into a integral
over the complex plane. This leads to 
\begin{align}\label{Eq_B17}
R_{II}(T) & =\frac{V_{pd}}{2}\sum\limits _{l,m=1}^{2}\int\biggl\{-\frac{1}{2\pi i}\oint_{\mathcal{C}_{1}}dzn_{F}(z)\biggl[\frac{1}{h_{l}^{(m)}(z,\mathbf{k})}\nonumber \\
 & \times\frac{\partial h_{l}^{(m)}(z,\mathbf{k})}{\partial R_{II}}+\frac{1}{\xoverline[0.8]{h}_{l}^{(m)}(z,\mathbf{k})}\frac{\partial\xoverline[0.8]{h}_{l}^{(m)}(z,\mathbf{k})}{\partial R_{II}}\biggr]\biggr\}\frac{d\mathbf{k}}{(2\pi)^{2}}.
\end{align}
Then we expand the terms between brackets in the above equation in
a series of partial fractions, i.e., 
\begin{align}\label{Eq_B18}
 & \frac{1}{h_{l}^{(m)}(z,\mathbf{k})}\frac{\partial h_{l}^{(m)}(z,\mathbf{k})}{\partial R_{II}}+\frac{1}{\xoverline[0.8]{h}_{l}^{(m)}(z,\mathbf{k})}\frac{\partial\xoverline[0.8]{h}_{l}^{(m)}(z,\mathbf{k})}{\partial R_{II}}\nonumber \\
 & =\sum\limits _{n=1}^{N}\dfrac{\Xi_{l,n}^{(m)}(\mathbf{k})}{z-\xi_{l,n}^{(m)}(\mathbf{k})},
\end{align}
where the coefficients of the expansion $\Xi_{l,n}^{(m)}(\mathbf{k})$
are given by 
\begin{align}
\Xi_{l,n}^{(m)}(\mathbf{k}) & =\dfrac{\dfrac{\partial h_{l}^{(m)}(z,\mathbf{k})}{\partial R_{II}}\bigg|_{z=\xi_{l,n}^{(m)}(\mathbf{k})}}{\dfrac{\partial h_{l}^{(m)}(z,\mathbf{k})}{\partial z}\bigg|_{z=\xi_{l,n}^{(m)}(\mathbf{k})}},\hspace{0.5cm}1\leqslant n\leqslant\frac{N}{2};\label{Eq_B19}\\
\Xi_{l,n}^{(m)}(\mathbf{k}) & =\dfrac{\dfrac{\partial\xoverline[0.8]{h}_{l}^{(m)}(z,\mathbf{k})}{\partial R_{II}}\bigg|_{z=\xi_{l,n}^{(m)}(\mathbf{k})}}{\dfrac{\partial\xoverline[0.8]{h}_{l}^{(m)}(z,\mathbf{k})}{\partial z}\bigg|_{z=\xi_{l,n}^{(m)}(\mathbf{k})}},\hspace{0.5cm}\frac{N}{2}+1\leqslant n\leqslant N.\label{Eq_B20}
\end{align}

Having in mind the result in Eq. \eqref{Eq_B18}, we evaluate the complex
integral in Eq. \eqref{Eq_B17} as 
\begin{align}\label{Eq_B21}
 & -\frac{1}{2\pi i}\sum\limits _{n=1}^{N}\Xi_{l,n}^{(m)}(\mathbf{k})\oint_{\mathcal{C}_{1}}dz\dfrac{n_{F}(z)}{z-\xi_{l,n}^{(m)}(\mathbf{k})}\nonumber \\
 & =\sum\limits _{n=1}^{N}\Xi_{l,n}^{(m)}(\mathbf{k})n_{F}[\xi_{l,n}^{(m)}(\mathbf{k})].
\end{align}

Lastly the mean-field equation for the loop-current order parameter
at finite temperature is obtained by inserting Eq. \eqref{Eq_B21}
into Eq. \eqref{Eq_B17}. Therefore this yields 
\begin{equation}\label{Eq_B22}
R_{II}(T)=\frac{V_{pd}}{2}\sum\limits _{l,m=1}^{2}\sum\limits _{n=1}^{N}\int\Xi_{l,n}^{(m)}(\mathbf{k})n_{F}[\xi_{l,n}^{(m)}(\mathbf{k})]\frac{d\mathbf{k}}{(2\pi)^{2}}.
\end{equation}
In the limit of zero temperature, this equation becomes 
\begin{equation}\label{Eq_B23}
R_{II}(T=0)=\frac{V_{pd}}{2}\sum\limits _{l,m=1}^{2}\sum\limits _{n=1}^{N}\int\Xi_{l,n}^{(m)}(\mathbf{k})\theta[-\xi_{l,n}^{(m)}(\mathbf{k})]\frac{d\mathbf{k}}{(2\pi)^{2}},
\end{equation}
with $\theta(-x)$ being the Fermi-Dirac distribution function $n_{F}(x)$
in this case.

\section{Form of the functions $\mathcal{D}_{l}^{(m)}(i\varepsilon_{n},\mathbf{k})$}

\noindent In order to compute the determinant $\text{det}[G^{-1}(i\varepsilon_{n},\mathbf{k})]$
that appears in the main text of this work, we need to make use of
the Eqs. \eqref{Eq_28} and \eqref{Eq_29}. Then as we are interested
in the interplay between loop-current and quadrupole density wave
orders, we also neglect the superconducting sector of the matrix $\hat{u}_{\tau}$.
As a result, this determinant evaluates to 
\begin{equation}\label{Eq_C1}
\det[G^{-1}(i\varepsilon_{n},\mathbf{k})]=\prod\limits _{l=1}^{2}\prod\limits _{m=1}^{2}\mathcal{D}_{l}^{(m)}(i\varepsilon_{n},\mathbf{k}),
\end{equation}
where $\mathcal{D}_{l}^{(m)}(i\varepsilon_{n},\mathbf{k})$ are functions
whose form will be determined in this appendix.

Before proceeding with that, let us define the following coefficients
\begin{align}
 & c_{1}(k_{x})=\sqrt{2}\left(-t_{pd}+i\frac{R_{II}}{4}k_{x}\right),\label{Eq_C2}\\
 & c_{1}(k_{y})=\sqrt{2}\left(-t_{pd}+i\frac{R_{II}}{4}k_{y}\right),\label{Eq_C3}\\
 & c_{2}(k_{x})=\frac{\sqrt{2}}{2}(-t_{pd}k_{x}-iR_{II}),\label{Eq_C4}\\
 & c_{2}(k_{y})=\frac{\sqrt{2}}{2}(-t_{pd}k_{y}-iR_{II}),\label{Eq_C5}
\end{align}
which are written as a function of Cu-O hopping $t_{pd}$, the loop-current
order parameter $R_{II}$, and the the momentum distance $\mathbf{k}$
to the hot spots. The purpose of defining these four coefficients
is of course to write down the functions $\mathcal{D}_{l}^{(m)}(i\varepsilon_{n},\mathbf{k})$
in a compact form.

We then construct the set of basis functions shown in Table \ref{TableI}
from the hot-spot parameter $\delta=(K_{+}-K_{-})/2$ and the $c_{i}(k_{x})$,
$c_{i}(k_{y})$ ($i=1,2$). As a consequence, we can write explicitly
the $\mathcal{D}_{l}^{(m)}(i\varepsilon_{n},\mathbf{k})$ as \begin{widetext}
\begin{align}
\mathcal{D}_{1}^{(1)}(i\varepsilon_{n},\mathbf{k}) & =|\{(-i\varepsilon_{n}+\xi_{d})[(-i\varepsilon_{n}+\xi_{p})^{2}-t_{pp}^{2}(a_{1}(\mathbf{k})+b_{1}(\mathbf{k}))]-P_{1}^{(0)}(\mathbf{k})(-i\varepsilon_{n}+\xi_{p})-t_{pp}P_{1}^{(1)}(\mathbf{k})\}\nonumber \\
 & \times\{(-i\varepsilon_{n}+\xi_{d})[(-i\varepsilon_{n}+\xi_{p})^{2}-t_{pp}^{2}(a_{2}(\mathbf{k})+b_{2}(\mathbf{k}))]-P_{2}^{(0)}(\mathbf{k})(-i\varepsilon_{n}+\xi_{p})-t_{pp}P_{2}^{(1)}(\mathbf{k})\}\nonumber \\
 & -b^{2}[(-i\varepsilon_{n}+\xi_{p})^{2}-t_{pp}^{2}(a_{1}(\mathbf{k})+b_{1}(\mathbf{k}))][(-i\varepsilon_{n}+\xi_{p})^{2}-t_{pp}^{2}(a_{2}(\mathbf{k})+b_{2}(\mathbf{k}))]|^{2},\label{Eq_C6}\\
\mathcal{D}_{1}^{(2)}(i\varepsilon_{n},\mathbf{k}) & =|\{(-i\varepsilon_{n}+\xi_{d})[(-i\varepsilon_{n}+\xi_{p})^{2}-t_{pp}^{2}(a_{1}(\mathbf{k})-b_{1}(\mathbf{k}))]-M_{1}^{(0)}(\mathbf{k})(-i\varepsilon_{n}+\xi_{p})-t_{pp}M_{1}^{(1)}(\mathbf{k})\}\nonumber \\
 & \times\{(-i\varepsilon_{n}+\xi_{d})[(-i\varepsilon_{n}+\xi_{p})^{2}-t_{pp}^{2}(a_{2}(\mathbf{k})-b_{2}(\mathbf{k}))]-M_{2}^{(0)}(\mathbf{k})(-i\varepsilon_{n}+\xi_{p})-t_{pp}M_{2}^{(1)}(\mathbf{k})\}\nonumber \\
 & -b^{2}[(-i\varepsilon_{n}+\xi_{p})^{2}-t_{pp}^{2}(a_{1}(\mathbf{k})-b_{1}(\mathbf{k}))][(-i\varepsilon_{n}+\xi_{p})^{2}-t_{pp}^{2}(a_{2}(\mathbf{k})-b_{2}(\mathbf{k}))]|^{2},\label{Eq_C7}\\
\mathcal{D}_{2}^{(1)}(i\varepsilon_{n},\mathbf{k}) & =|\{(-i\varepsilon_{n}+\xi_{d})[(-i\varepsilon_{n}+\xi_{p})^{2}-t_{pp}^{2}(a_{3}(\mathbf{k})+b_{3}(\mathbf{k}))]-P_{3}^{(0)}(\mathbf{k})(-i\varepsilon_{n}+\xi_{p})-t_{pp}P_{3}^{(1)}(\mathbf{k})\}\nonumber\\
 & \times\{(-i\varepsilon_{n}+\xi_{d})[(-i\varepsilon_{n}+\xi_{p})^{2}-t_{pp}^{2}(a_{4}(\mathbf{k})+b_{4}(\mathbf{k}))]-P_{4}^{(0)}(\mathbf{k})(-i\varepsilon_{n}+\xi_{p})-t_{pp}P_{4}^{(1)}(\mathbf{k})\}\nonumber\\
 & -b^{2}[(-i\varepsilon_{n}+\xi_{p})^{2}-t_{pp}^{2}(a_{3}(\mathbf{k})+b_{3}(\mathbf{k}))][(-i\varepsilon_{n}+\xi_{p})^{2}-t_{pp}^{2}(a_{4}(\mathbf{k})+b_{4}(\mathbf{k}))]|^{2},\label{Eq_C8}\\
\mathcal{D}_{2}^{(2)}(i\varepsilon_{n},\mathbf{k}) & =|\{(-i\varepsilon_{n}+\xi_{d})[(-i\varepsilon_{n}+\xi_{p})^{2}-t_{pp}^{2}(a_{3}(\mathbf{k})-b_{3}(\mathbf{k}))]-M_{3}^{(0)}(\mathbf{k})(-i\varepsilon_{n}+\xi_{p})-t_{pp}M_{3}^{(1)}(\mathbf{k})\}\nonumber \\
 & \times\{(-i\varepsilon_{n}+\xi_{d})[(-i\varepsilon_{n}+\xi_{p})^{2}-t_{pp}^{2}(a_{4}(\mathbf{k})-b_{4}(\mathbf{k}))]-M_{4}^{(0)}(\mathbf{k})(-i\varepsilon_{n}+\xi_{p})-t_{pp}M_{4}^{(1)}(\mathbf{k})\}\nonumber \\
 & -b^{2}[(-i\varepsilon_{n}+\xi_{p})^{2}-t_{pp}^{2}(a_{3}(\mathbf{k})-b_{3}(\mathbf{k}))][(-i\varepsilon_{n}+\xi_{p})^{2}-t_{pp}^{2}(a_{4}(\mathbf{k})-b_{4}(\mathbf{k}))]|^{2},\label{Eq_C9}
\end{align}
\end{widetext} where we have used a second set of basis functions
defined in Table \ref{TableII} as well as the new functions 
\begin{align}
P_{l}^{(0)}(\mathbf{k}) & =a_{lx}(k_{x})+a_{ly}(k_{y})+b_{lx}(k_{x})+b_{ly}(k_{y}),\label{Eq_C10}\\
P_{l}^{(1)}(\mathbf{k}) & =[\widetilde{a}_{l}(\mathbf{k})+\widetilde{b}_{l}(\mathbf{k})][a_{lxy}(\mathbf{k})+b_{lxy}(\mathbf{k})],\label{Eq_C11}\\
M_{l}^{(0)}(\mathbf{k}) & =a_{lx}(k_{x})+a_{ly}(k_{y})-b_{lx}(k_{x})-b_{ly}(k_{y}),\label{Eq_C12}\\
M_{l}^{(1)}(\mathbf{k}) & =[\widetilde{a}_{l}(\mathbf{k})-\widetilde{b}_{l}(\mathbf{k})][a_{lxy}(\mathbf{k})-b_{lxy}(\mathbf{k})],\label{Eq_C13}
\end{align}
which depend on the basis functions of both tables.

\begin{table*}
\centering
\caption{First set of basis functions used to represent the free energy of the three-band model. Here these functions are written in terms of the hot-spot parameter $\delta=(K_{+}-K_{-})/2$ and the coefficients $c_{i}(k_{x})$ and $c_{i}(k_{y})$ ($i=1,2$).}\label{TableI}
\begin{tabular}{cccccc}
\hline
\hline
Basis function         & Definition \\ \hline
$a_{1x}(k_{x})$  &  $|c_{1}(k_{x})|^{2}+|c_{2}(k_{x})|^{2}$   \\
$a_{2x}(k_{x})$  &  $|c_{1}(k_{x})|^{2}+|c_{2}(k_{x})|^{2}$   \\
$a_{3x}(k_{x})$  &  $|c_{1}(k_{x})|^{2}+|c_{2}(k_{x})|^{2}$   \\
$a_{4x}(k_{x})$  &  $|c_{1}(k_{x})|^{2}+|c_{2}(k_{x})|^{2}$   \\
$b_{1x}(k_{x})$  &  $\sin\delta[|c_{1}(k_{x})|^{2}-|c_{2}(k_{x})|^{2}]+2\cos\delta\operatorname{Re}[c^{*}_{1}(k_{x})c_{2}(k_{x})]$   \\
$b_{2x}(k_{x})$  &  $\sin\delta[|c_{1}(k_{x})|^{2}-|c_{2}(k_{x})|^{2}]-2\cos\delta\operatorname{Re}[c^{*}_{1}(k_{x})c_{2}(k_{x})]$   \\
$b_{3x}(k_{x})$  &  $\sin\delta[|c_{2}(k_{x})|^{2}-|c_{1}(k_{x})|^{2}]+2\cos\delta\operatorname{Re}[c^{*}_{1}(k_{x})c_{2}(k_{x})]$   \\
$b_{4x}(k_{x})$  &  $\sin\delta[|c_{2}(k_{x})|^{2}-|c_{1}(k_{x})|^{2}]-2\cos\delta\operatorname{Re}[c^{*}_{1}(k_{x})c_{2}(k_{x})]$   \\
$a_{1y}(k_{y})$  &  $|c_{1}(k_{y})|^{2}+|c_{2}(k_{y})|^{2}$   \\
$a_{2y}(k_{y})$  &  $|c_{1}(k_{y})|^{2}+|c_{2}(k_{y})|^{2}$   \\
$a_{3y}(k_{y})$  &  $|c_{1}(k_{y})|^{2}+|c_{2}(k_{y})|^{2}$   \\
$a_{4y}(k_{y})$  &  $|c_{1}(k_{y})|^{2}+|c_{2}(k_{y})|^{2}$   \\
$b_{1y}(k_{y})$  &  $\sin\delta[|c_{2}(k_{y})|^{2}-|c_{1}(k_{y})|^{2}]+2\cos\delta\operatorname{Re}[c^{*}_{1}(k_{y})c_{2}(k_{y})]$   \\
$b_{2y}(k_{y})$  &  $\sin\delta[|c_{2}(k_{y})|^{2}-|c_{1}(k_{y})|^{2}]-2\cos\delta\operatorname{Re}[c^{*}_{1}(k_{y})c_{2}(k_{y})]$   \\
$b_{3y}(k_{y})$  &  $\sin\delta[|c_{1}(k_{y})|^{2}-|c_{2}(k_{y})|^{2}]-2\cos\delta\operatorname{Re}[c^{*}_{1}(k_{y})c_{2}(k_{y})]$   \\
$b_{4y}(k_{y})$  &  $\sin\delta[|c_{1}(k_{y})|^{2}-|c_{2}(k_{y})|^{2}]+2\cos\delta\operatorname{Re}[c^{*}_{1}(k_{y})c_{2}(k_{y})]$   \\
$a_{1xy}(\mathbf{k})$  &  $2\cos\delta\operatorname{Re}[c^{*}_{1}(k_{x})c_{1}(k_{y})+c^{*}_{2}(k_{x})c_{2}(k_{y})]+2\sin\delta\operatorname{Re}[c^{*}_{1}(k_{x})c_{2}(k_{y})-c^{*}_{2}(k_{x})c_{1}(k_{y})]$   \\
$a_{2xy}(\mathbf{k})$  &  $2\cos\delta\operatorname{Re}[c^{*}_{1}(k_{x})c_{1}(k_{y})+c^{*}_{2}(k_{x})c_{2}(k_{y})]-2\sin\delta\operatorname{Re}[c^{*}_{1}(k_{x})c_{2}(k_{y})-c^{*}_{2}(k_{x})c_{1}(k_{y})]$   \\
$a_{3xy}(\mathbf{k})$  &  $2\cos\delta\operatorname{Re}[c^{*}_{1}(k_{x})c_{1}(k_{y})-c^{*}_{2}(k_{x})c_{2}(k_{y})]+2\sin\delta\operatorname{Re}[c^{*}_{1}(k_{x})c_{2}(k_{y})+c^{*}_{2}(k_{x})c_{1}(k_{y})]$   \\
$a_{4xy}(\mathbf{k})$  &  $2\cos\delta\operatorname{Re}[c^{*}_{1}(k_{x})c_{1}(k_{y})-c^{*}_{2}(k_{x})c_{2}(k_{y})]-2\sin\delta\operatorname{Re}[c^{*}_{1}(k_{x})c_{2}(k_{y})+c^{*}_{2}(k_{x})c_{1}(k_{y})]$   \\
$b_{1xy}(\mathbf{k})$  &  $2\operatorname{Re}[c^{*}_{2}(k_{x})c_{1}(k_{y})+c^{*}_{1}(k_{x})c_{2}(k_{y})]$   \\
$b_{2xy}(\mathbf{k})$  & $-2\operatorname{Re}[c^{*}_{2}(k_{x})c_{1}(k_{y})+c^{*}_{1}(k_{x})c_{2}(k_{y})]$    \\
$b_{3xy}(\mathbf{k})$  &  $2\operatorname{Re}[c^{*}_{2}(k_{x})c_{1}(k_{y})-c^{*}_{1}(k_{x})c_{2}(k_{y})]$   \\
$b_{4xy}(\mathbf{k})$  &  $-2\operatorname{Re}[c^{*}_{2}(k_{x})c_{1}(k_{y})-c^{*}_{1}(k_{x})c_{2}(k_{y})]$   \\
\hline
\hline
\end{tabular}
\end{table*}

\begin{table*}
\centering
\caption{Second set of basis functions needed to evaluate the free energy of the present three-band model. In our notation, the indices $l$ and $\widetilde{l}$ refer respectively to the functions $a_{l}(\mathbf{k})$ [and $b_{l}(\mathbf{k})$] and $\widetilde{a}_{l}(\mathbf{k})$ [and $\widetilde{b}_{l}(\mathbf{k})$].}\label{TableII}
\begin{center}
\begin{tabular}{cccccc}
\hline
\hline
$l$         & \ & $a_{l}(\mathbf{k})$ & \ & $b_{l}(\mathbf{k})$ \\ \hline
$1$  & \ &  $(k_{x}+k_{y})^{2}+\sin^{2}\delta(k_{y}-k_{x}+2\cot\delta)^{2}$  & \ &  $2\sin\delta[(k_{y}+\cot\delta)^{2}-(k_{x}-\cot\delta)^{2}]$   \\
$2$  & \ &  $(k_{x}+k_{y})^{2}+\sin^{2}\delta(k_{y}-k_{x}-2\cot\delta)^{2}$  & \ &  $2\sin\delta[(k_{y}-\cot\delta)^{2}-(k_{x}+\cot\delta)^{2}]$   \\
$3$  & \ &  $(k_{x}-k_{y})^{2}+\sin^{2}\delta(k_{x}+k_{y}+2\cot\delta)^{2}$  & \ &  $2\sin\delta[(k_{x}+\cot\delta)^{2}-(k_{y}+\cot\delta)^{2}]$   \\
$4$  & \ &  $(k_{x}-k_{y})^{2}+\sin^{2}\delta(k_{x}+k_{y}-2\cot\delta)^{2}$  & \ &  $2\sin\delta[(k_{x}-\cot\delta)^{2}-(k_{y}-\cot\delta)^{2}]$   \\
$\widetilde{1}$  & \ &  $-\sin\delta(k_{x}-k_{y}-2\cot\delta)$  & \ &  $k_{x}+k_{y}$   \\
$\widetilde{2}$  & \ &  $\sin\delta(k_{x}-k_{y}+2\cot\delta)$  & \ &  $-(k_{x}+k_{y})$   \\
$\widetilde{3}$  & \ &  $\sin\delta(k_{x}+k_{y}+2\cot\delta)$  & \ &  $k_{x}-k_{y}$   \\
$\widetilde{4}$  & \ &  $-\sin\delta(k_{x}+k_{y}-2\cot\delta)$  & \ &  $-(k_{x}-k_{y})$   \\
\hline
\hline
\end{tabular}
\end{center}
\end{table*}

In the main text of this paper, we have to perform the analytic continuation
$\varepsilon_{n}\rightarrow-iz$ for $\mathcal{D}_{l}^{(m)}(i\varepsilon_{n},\mathbf{k})$.
By observing the results in Eqs. \eqref{Eq_C6}--\eqref{Eq_C9}, we
conclude that each $\mathcal{D}_{l}^{(m)}(i\varepsilon_{n},\mathbf{k})$
could be written as a product of a function times its complex conjugate.
Therefore we can make the analytic continuation as follows 
\begin{equation}\label{Eq_C14}
\mathcal{D}_{l}^{(m)}(z,\mathbf{k})=h_{l}^{(m)}(z,\mathbf{k})\xoverline[0.8]{h}_{l}^{(m)}(z,\mathbf{k}),
\end{equation}
where the functions on the right-hand side of the above equality are
given by \begin{widetext} 
\begin{align}
h_{1}^{(1)}(z,\mathbf{k}) & =\{(z-\xi_{d})[(z-\xi_{p})^{2}-t_{pp}^{2}(a_{1}(\mathbf{k})+b_{1}(\mathbf{k}))]-P_{1}^{(0)}(\mathbf{k})(z-\xi_{p})+t_{pp}P_{1}^{(1)}(\mathbf{k})\}\nonumber \\
 & \times\{(z-\xi_{d})[(z-\xi_{p})^{2}-t_{pp}^{2}(a_{2}(\mathbf{k})+b_{2}(\mathbf{k}))]-P_{2}^{(0)}(\mathbf{k})(z-\xi_{p})+t_{pp}P_{2}^{(1)}(\mathbf{k})\}\nonumber \\
 & -b^{2}[(z-\xi_{p})^{2}-t_{pp}^{2}(a_{1}(\mathbf{k})+b_{1}(\mathbf{k}))][(z-\xi_{p})^{2}-t_{pp}^{2}(a_{2}(\mathbf{k})+b_{2}(\mathbf{k}))],\label{Eq_C15}\\
\xoverline[0.8]{h}_{1}^{(1)}(z,\mathbf{k}) & =\{(z+\xi_{d})[(z+\xi_{p})^{2}-t_{pp}^{2}(a_{1}(\mathbf{k})+b_{1}(\mathbf{k}))]-P_{1}^{(0)}(\mathbf{k})(z+\xi_{p})-t_{pp}P_{1}^{(1)}(\mathbf{k})\}\nonumber \\
 & \times\{(z+\xi_{d})[(z+\xi_{p})^{2}-t_{pp}^{2}(a_{2}(\mathbf{k})+b_{2}(\mathbf{k}))]-P_{2}^{(0)}(\mathbf{k})(z+\xi_{p})-t_{pp}P_{2}^{(1)}(\mathbf{k})\}\nonumber \\
 & -b^{2}[(z+\xi_{p})^{2}-t_{pp}^{2}(a_{1}(\mathbf{k})+b_{1}(\mathbf{k}))][(z+\xi_{p})^{2}-t_{pp}^{2}(a_{2}(\mathbf{k})+b_{2}(\mathbf{k}))],\label{Eq_C16}\\
h_{1}^{(2)}(z,\mathbf{k}) & =\{(z-\xi_{d})[(z-\xi_{p})^{2}-t_{pp}^{2}(a_{1}(\mathbf{k})-b_{1}(\mathbf{k}))]-M_{1}^{(0)}(\mathbf{k})(z-\xi_{p})+t_{pp}M_{1}^{(1)}(\mathbf{k})\}\nonumber \\
 & \times\{(z-\xi_{d})[(z+\xi_{p})^{2}-t_{pp}^{2}(a_{2}(\mathbf{k})-b_{2}(\mathbf{k}))]-M_{2}^{(0)}(\mathbf{k})(z-\xi_{p})+t_{pp}M_{2}^{(1)}(\mathbf{k})\}\nonumber \\
 & -b^{2}[(z-\xi_{p})^{2}-t_{pp}^{2}(a_{1}(\mathbf{k})-b_{1}(\mathbf{k}))][(z-\xi_{p})^{2}-t_{pp}^{2}(a_{2}(\mathbf{k})-b_{2}(\mathbf{k}))],\label{Eq_C17}
\end{align}
\begin{align}
\xoverline[0.8]{h}_{1}^{(2)}(z,\mathbf{k}) & =\{(z+\xi_{d})[(z+\xi_{p})^{2}-t_{pp}^{2}(a_{1}(\mathbf{k})-b_{1}(\mathbf{k}))]-M_{1}^{(0)}(\mathbf{k})(z+\xi_{p})-t_{pp}M_{1}^{(1)}(\mathbf{k})\}\nonumber \\
 & \times\{(z+\xi_{d})[(z+\xi_{p})^{2}-t_{pp}^{2}(a_{2}(\mathbf{k})-b_{2}(\mathbf{k}))]-M_{2}^{(0)}(\mathbf{k})(z+\xi_{p})-t_{pp}M_{2}^{(1)}(\mathbf{k})\}\nonumber \\
 & -b^{2}[(z+\xi_{p})^{2}-t_{pp}^{2}(a_{1}(\mathbf{k})-b_{1}(\mathbf{k}))][(z+\xi_{p})^{2}-t_{pp}^{2}(a_{2}(\mathbf{k})-b_{2}(\mathbf{k}))],\label{Eq_C18}\\
h_{2}^{(1)}(z,\mathbf{k}) & =\{(z-\xi_{d})[(z-\xi_{p})^{2}-t_{pp}^{2}(a_{3}(\mathbf{k})+b_{3}(\mathbf{k}))]-P_{3}^{(0)}(\mathbf{k})(z-\xi_{p})+t_{pp}P_{3}^{(1)}(\mathbf{k})\}\nonumber \\
 & \times\{(z-\xi_{d})[(z-\xi_{p})^{2}-t_{pp}^{2}(a_{4}(\mathbf{k})+b_{4}(\mathbf{k}))]-P_{4}^{(0)}(\mathbf{k})(z-\xi_{p})+t_{pp}P_{4}^{(1)}(\mathbf{k})\}\nonumber \\
 & -b^{2}[(z-\xi_{p})^{2}-t_{pp}^{2}(a_{3}(\mathbf{k})+b_{3}(\mathbf{k}))][(z-\xi_{p})^{2}-t_{pp}^{2}(a_{4}(\mathbf{k})+b_{4}(\mathbf{k}))],\label{Eq_C19}\\
\xoverline[0.8]{h}_{2}^{(1)}(z,\mathbf{k}) & =\{(z+\xi_{d})[(z+\xi_{p})^{2}-t_{pp}^{2}(a_{3}(\mathbf{k})+b_{3}(\mathbf{k}))]-P_{3}^{(0)}(\mathbf{k})(z+\xi_{p})-t_{pp}P_{3}^{(1)}(\mathbf{k})\}\nonumber \\
 & \times\{(z+\xi_{d})[(z+\xi_{p})^{2}-t_{pp}^{2}(a_{4}(\mathbf{k})+b_{4}(\mathbf{k}))]-P_{4}^{(0)}(\mathbf{k})(z+\xi_{p})-t_{pp}P_{4}^{(1)}(\mathbf{k})\}\nonumber \\
 & -b^{2}[(z+\xi_{p})^{2}-t_{pp}^{2}(a_{3}(\mathbf{k})+b_{3}(\mathbf{k}))][(z+\xi_{p})^{2}-t_{pp}^{2}(a_{4}(\mathbf{k})+b_{4}(\mathbf{k}))],\label{Eq_C20}
\end{align}
\begin{align}
h_{2}^{(2)}(z,\mathbf{k}) & =\{(z-\xi_{d})[(z-\xi_{p})^{2}-t_{pp}^{2}(a_{3}(\mathbf{k})-b_{3}(\mathbf{k}))]-M_{3}^{(0)}(\mathbf{k})(z-\xi_{p})+t_{pp}M_{3}^{(1)}(\mathbf{k})\}\nonumber \\
 & \times\{(z-\xi_{d})[(z+\xi_{p})^{2}-t_{pp}^{2}(a_{4}(\mathbf{k})-b_{4}(\mathbf{k}))]-M_{4}^{(0)}(\mathbf{k})(z-\xi_{p})+t_{pp}M_{4}^{(1)}(\mathbf{k})\}\nonumber \\
 & -b^{2}[(z-\xi_{p})^{2}-t_{pp}^{2}(a_{3}(\mathbf{k})-b_{3}(\mathbf{k}))][(z-\xi_{p})^{2}-t_{pp}^{2}(a_{4}(\mathbf{k})-b_{4}(\mathbf{k}))],\label{Eq_C21}\\
\xoverline[0.8]{h}_{2}^{(2)}(z,\mathbf{k}) & =\{(z+\xi_{d})[(z+\xi_{p})^{2}-t_{pp}^{2}(a_{3}(\mathbf{k})-b_{3}(\mathbf{k}))]-M_{3}^{(0)}(\mathbf{k})(z+\xi_{p})-t_{pp}M_{3}^{(1)}(\mathbf{k})\}\nonumber \\
 & \times\{(z+\xi_{d})[(z+\xi_{p})^{2}-t_{pp}^{2}(a_{4}(\mathbf{k})-b_{4}(\mathbf{k}))]-M_{4}^{(0)}(\mathbf{k})(z+\xi_{p})-t_{pp}M_{4}^{(1)}(\mathbf{k})\}\nonumber \\
 & -b^{2}[(z+\xi_{p})^{2}-t_{pp}^{2}(a_{3}(\mathbf{k})-b_{3}(\mathbf{k}))][(z+\xi_{p})^{2}-t_{pp}^{2}(a_{4}(\mathbf{k})-b_{4}(\mathbf{k}))].\label{Eq_C22}
\end{align}
\end{widetext}
According to the approach developed in the Appendix
B for solving the LC and QDW mean-field equations, we need to determine
first the roots of $h_{l}^{(m)}(z,\mathbf{k})$ and $\xoverline[0.8]{h}_{l}^{(m)}(z,\mathbf{k})$
which are denoted here as $\xi_{l,n}^{(m)}(\mathbf{k})$ $(n=1,\ldots,N)$.
As $h_{l}^{(m)}(z,\mathbf{k})$ and $\xoverline[0.8]{h}_{l}^{(m)}(z,\mathbf{k})$
are both sixth-order polynomials in the variable $z$, their roots
will be determined by means of numerical methods.

\end{document}